\documentstyle[12pt]{article}
\begin{document}
\title{Ground-state properties\\
of deformed proton emitters\\
in the relativistic Hartree-Bogoliubov model} 
\author{G.A. Lalazissis$^{1,3}$, D. Vretenar$^{1,2}$, and P. Ring$^{1}$
\vspace{0.5 cm}\\
$^{1}$ Physik-Department der Technischen Universit\"at M\"unchen,\\
D-85748 Garching, Germany\\
$^{2}$ Physics Department, Faculty of Science, 
\\University of Zagreb, 1000 Zagreb, Croatia\\
$^{3}$ National Superconducting Cyclotron Laboratory, \\
Michigan State University
East Lansing, Michigan 48825, USA}
\maketitle
\bigskip
\bigskip
\begin{abstract}
The Relativistic Hartree Bogoliubov (RHB) model is applied in the
description of ground-state properties of proton-rich odd-Z
nuclei in the region $53 \leq Z \leq 69$. The NL3 effective
interaction is used in the mean-field Lagrangian,
and pairing correlations are described by the 
pairing part of the finite range Gogny interaction D1S.
The model predicts the location of the proton drip-line,
the ground-state quadrupole deformations and 
one-proton separation energies at and beyond the drip-line,
the deformed single-particle 
orbitals occupied by the odd valence proton, and the 
corresponding spectroscopic factors. The results of 
fully self-consistent RHB calculations are compared 
with available experimental data, and with predictions
of the macroscopic-microscopic mass model.
\end{abstract}
\bigskip \bigskip

\vspace{1 cm} {PACS numbers:} {21.60.Jz, 21.10.Dr, 21.10.Jx, 23.50.+z,
27.60+j}\newline
\vspace{1 cm}\newline
\newpage
\baselineskip = 24pt

%
\section{Introduction}
Models based on the relativistic mean-field approximation provide a
microscopically consistent, and yet simple and economical description of the
nuclear many-body problem. By adjusting just a few model parameters:
coupling constants and effective masses, to global properties of simple,
spherical and stable nuclei, it has been possible to describe in detail a
variety of nuclear structure phenomena over the whole periodic table, from
light nuclei to superheavy elements \cite{Rin.96}. When also pairing
correlations are included in the self-consistent Hartree-Bogoliubov
framework, the relativistic mean-field theory can be applied to the physics
of exotic nuclei at the drip-lines.

The relativistic Hartree-Bogoliubov (RHB) model represents a relativistic
extension of the Hartree-Fock-Bogoliubov (HFB) framework. A unified and
self-consistent description of mean-field and pairing correlations is
especially important in applications to drip-line nuclei. 
At the drip-lines the
separation energy of the last nucleons becomes extremely small or 
negative, the Fermi level is found close to particle continuum, and the
lowest particle-hole or particle-particle modes are embedded in the
continuum. The RHB model with finite range pairing interactions has been
applied in studies of the halo phenomenon in light nuclei 
\cite{MR.96,PVL.97},
properties of light nuclei near the neutron-drip \cite{LVP.98}, 
reduction of the spin-orbit potential in nuclei with extreme isospin values 
\cite{LVR.97}, ground-state properties of Ni and Sn isotopes 
\cite{LVR.98}, and the deformation and shape coexistence 
phenomena that result from the
suppression of the spherical N=28 shell gap in neutron-rich nuclei 
\cite{LVR.98a}. In particular, in Ref. \cite{VLR.98} we have applied the RHB
model to calculate properties of proton-rich spherical even-even nuclei with
14$\leq $Z$\leq $28 and N=18,20,22. It has been shown that RHB provides a
very good description of binding energies, two-proton separation energies
and proton ${\it rms}$ radii. Model predictions for the proton drip-line
have been found in excellent agreement with shell-model calculations 
and with results of non-relativistic HF and HFB
studies. The position of the two-proton drip-line for even-even nuclei with
10$\leq $Z$\leq $82 has been recently determined with the RMF+BCS model 
\cite{LR.98}. Results computed with the NL3 effective interaction 
\cite{LKR.97}
have been compared with those of the Hartree-Fock+BCS model (effective force
Skyrme SIII) \cite{TTO.96}, the finite-range droplet model (FRDM) 
\cite{MN.95,MNK.97}, and with the available experimental information. In a very
recent work \cite{Vre.98L} we have used the RHB model with finite range
pairing for an analysis of the structure of deformed proton-rich odd-Z
nuclei in the region 59$\leq $Z$\leq $69. The location of the proton
drip-line, the properties of proton emitters beyond the drip-line, and the
deformed single-particle orbitals occupied by the odd valence proton have
been compared with available experimental data.

The structure of proton-rich nuclei displays many interesting properties
which are important both for nuclear physics and astrophysics. These nuclei
are characterized by exotic decay modes, such as the direct emission of charged
particles from the ground-state, and $\beta$-decay with large Q-values.
The properties of most proton-rich nuclei should also play an important role
in the process of nucleosynthesis by rapid-proton capture. The decay by
direct proton emission provides the opportunity to study the structure of
systems beyond the drip-line. Proton radioactivity from the ground-state is
determined by the Coulomb and centrifugal terms of the effective potential.
For Z$\leq $50, nuclei beyond the proton drip-line exist only as short lived
resonances, and ground-state proton decay probably cannot be observed
directly \cite{WD.97}. On the other hand, 
the relatively high potential energy barrier
enables the observation of ground-state proton emission from medium-heavy
and heavy nuclei. At the drip-lines proton emission competes with 
$\beta^{+}$ decay; for heavy nuclei also fission or $\alpha $ decay can be
favored. The proton-drip line has been fully mapped up to Z=21, and
possibly for odd-Z nuclei up to In \cite{WD.97}. Detailed studies of
ground-state proton radioactivity have been reported for odd-Z nuclei in the
two spherical regions 51$\leq $Z$\leq $55 and 69$\leq $Z$\leq $ 83. The
systematics of spectroscopic factors is consistent with
half-lives calculated in the spherical WKB or DWBA approximations
\cite{Sel.93,Dav.97,ASN.97}. Recently
reported proton decay rates \cite{Dav.98} indicate that the missing region
of light rare-earth nuclei contains strongly deformed systems at the
drip-lines. The lifetimes of deformed proton emitters should provide direct
information on the Nilsson configuration occupied by the odd proton, and
therefore on the shape of the nucleus.

In the present work we extended the analysis of Ref. \cite{Vre.98L}, and
report a detailed study of deformed odd-Z proton-rich nuclei with 
53$\leq$Z$\leq$69. In Sec. 2 we present an outline of the relativistic
Hartree-Bogoliubov theory and discuss the various approximations and
symmetry constraints. In Sec. 3 the model is applied in the calculation of
ground-state properties of proton-rich isotopes at and beyond the drip-line.
The results are summarized in Sec. 4.

\bigskip
%

\section{Relativistic Hartree Bogoliubov model with finite range pairing
interaction}


\subsection{The Relativistic Mean Field Model}

The relativistic mean field theory presents
a phenomenological description of the nuclear system. 
The theory is based on very simple concepts:
(i) nucleons are described as point particles, 
(ii) these particles strictly obey the rules 
of relativity and causality, and (iii) they move independently in
mean fields which originate from the nucleon-nucleon interaction. 
Conditions of causality and Lorentz invariance impose that the 
interaction is mediated by the
exchange of point-like effective mesons, which couple to the nucleons  
at local vertices. The model parameters are the meson
masses and their coupling constants. The values of the
parameters are adjusted to properties of nuclear
matter and experimental data on finite nuclear systems.  

On this phenomenological level the different mesons are classified
according to the quantum numbers of spin, parity and isospin. 
The concepts of simplicity, and the desire to keep the number of 
phenomenological parameters as small as possible, impose that
only the most important meson fields should be included in 
the model. The experimental fact that parity is conserved
to a very high degree of accuracy, on the mean field level 
automatically excludes mesons with negative intrinsic parity. 
Since the model describes a system with
two types of particles: neutrons and protons, it is necessary 
to introduce both isoscalar and isovector meson fields.
Most versions of the relativistic mean-field model, however,
do not contain the scalar isovector meson $\delta $. From the presently 
available experimental data it is difficult to determine its 
effective coupling constant and effective mass, i.e. the same 
level accuracy is obtained with or without the inclusion
of the $\delta $-meson. On the other hand, essential for a
correct description of properties of finite nuclei is the 
vector isovector $\vec \rho $-meson. 
 
The nucleons are described as relativistic particles moving
independently in local mean-field meson potentials. The single-nucleon
relativistic dynamics is described by the Dirac equation. In
contrast to the Schr\"{o}dinger equation, which contains a single
mean-field potential of Woods-Saxon shape and an
independently adjusted spin-orbit potential, 
the Lorentz structure of the Dirac equation distinguishes between the
scalar potential $S({\bf r})$, which results from 
the exchange of spin-zero mesons, 
and a vector potential induced by spin-one mesons. In addition, the 
spin-orbit potential is automatically included in the equation of 
motion, i.e. it does not require an independent strength 
parameter. The vector potential 
$(V_{0}({\bf r),V(r))}$ is a four-vector under Lorentz transformations. 
It is similar in structure to the electromagnetic potentials 
$(A_{0}({\bf r),A(r))}$ which occur when the Dirac equation is applied
to systems in atomic physics. The time-like component of
the vector field $V_{0}({\bf r)}$ corresponds to the Coulomb field 
$A_{0}({\bf r})$, and the three space-like components ${\bf V(r)}$ are
equivalent to the magnetic potential ${\bf A}({\bf r})$.

In calculations of properties of ground-state proton emitters we neglect
the space-like components of the vector fields,
i.e. we neglect nuclear magnetism. For nuclei with even number of protons  
and neutrons this is certainly justified by the time-reversal
invariance of the mean field. In nuclei with an odd number of protons or
neutrons, time-reversal symmetry is broken. The odd particle induces
polarization currents and time-odd components in the meson fields. The
time-odd components are essential for a description of magnetic moments
\cite{HR.88}, and of moments of inertia in rotating nuclei 
\cite{KR.93,AKR.96}. The effect on deformations
and binding energies, however, is very small and can be neglected to a good
approximation. We are then left with the
time-like component $V({\bf r)}$ (for simplicity we omit the subscript $0$
henceforth). The Lorentz structure of the theory implies that the scalar field
is attractive, and that the time-like component of the vector 
fields is repulsive. Hence, the scalar field can be used to characterize
in an economic way the attractive part
of the nucleon-nucleon interaction at intermediate distances, and the vector
fields describe the repulsion at short distances.
 
Under these simplifying assumptions the single-nucleon dynamics is 
described by the Dirac equation, which 
contains only the time-like component of the vector potential 
$V$ and the scalar potential $S$. In the stationary case it 
reduces to an eigenvalue problem 
\begin{equation}
\hat h_D \psi_i =
\{{\bf \alpha p}+V({\bf r})+\beta \lbrack m-S({\bf r})]\}~\psi
_{i}=~\epsilon _{i}\psi _{i}.  
\label{dirac1}
\end{equation}
This equation contains the four-dimensional Dirac matrices ${\bf \alpha }$
and $\beta $, and the nucleons with rest mass $m$ are represented by the
four-dimensional vectors $\psi_i$. Expressed in
terms of the Pauli spin matrices, the equation reads 
\begin{equation}
\left( 
\begin{array}{cc}
m-S+V & {\bf \sigma p} \\ 
{\bf \sigma p} & -m+S+V
\end{array}
\right) \left( 
\begin{array}{l}
f \\ 
g
\end{array}
\right) _{i}=\varepsilon _{i}\left( 
\begin{array}{l}
f \\ 
g
\end{array}
\right) _{i}.  
\label{dirac2}
\end{equation}
The single-particle wave functions $\psi _{i}$ are four-dimensional spinors.
The subscript $i$ denotes the quantum numbers that specify the 
single-particle state, and $\varepsilon _{i}$ is the corresponding energy.
In the {\it no-sea approximation}
the Dirac sea of solutions with negative eigenvalues does not 
contribute to the densities and currents. Performing only the summation
over the occupied orbitals in the Fermi sea, the bilinear
products of wave functions are used to construct 
two types of densities: (i) the baryon density 
\begin{equation}
\rho ({\bf r)=}\sum_{i=1}^{A}\psi _{i}^{+}\psi
_{i}^{{}}=\sum_{i=1}^{A}[f_{i}^{+}({\bf r)}f_{i}^{{}}({\bf r)}+g_{i}^{+}({\bf %
r)}g_{i}^{{}}({\bf r)],}  \label{rhov}
\end{equation}
which is the zero component of the four-vector baryon current, 
and (ii) the scalar density 
\begin{equation}
\rho _{s}({\bf r)=}\sum_{i=1}^{A}\overline{\psi }_{i}\psi
_{i}^{{}}=\sum_{i=1}^{A}[f_{i}^{+}({\bf r)}f_{i}^{{}}({\bf r)}-g_{i}^{+}({\bf %
r)}g_{i}^{{}}({\bf r)].}  \label{rhos}
\end{equation}
The fields $V({\bf r})$ and 
$S({\bf r})$ are computed by averaging over the interactions induced by the
exchange of vector and scalar mesons, respectively
\begin{eqnarray}
V({\bf r}) &=&\int v_{v}({\bf r,r}^{\prime })\rho ({\bf r}^{\prime })d^{3}r,
\label{v1} \\
S({\bf r}) &=&\int v_{s}({\bf r,r}^{\prime })\rho _{s}({\bf r}^{\prime
})d^{3}r.  \label{s1}
\end{eqnarray}
The two-body interactions $v_{s}({\bf r,r}^{\prime })$ and 
$v_{v}({\bf r,r}^{\prime })$ are of Yukawa type. 
They correspond to the exchange of the scalar
isoscalar $\sigma $-meson, and vector mesons: $\omega $
(isoscalar) and $\vec{\rho}$ (isovector). The resulting fields are defined  
\begin{eqnarray}
S({\bf r}) &=&-g_{\sigma }\sigma ({\bf r}),  \label{v2} \\
V({\bf r}) &=&g_{\omega }\omega ({\bf r})+g_{\rho }\rho _{3}({\bf r})+A_{0}(%
{\bf r}),  \label{s2}
\end{eqnarray}
where $\sigma ({\bf r})$, $\omega ({\bf r})$, $\rho _{3}({\bf r})$ are
the classical mesons fields, and $A_{0}({\bf r})$ is the Coulomb
field (exchange of photons). 
The mesons fields are solutions of the Klein-Gordon equations\smallskip 
\begin{eqnarray}
(-\Delta +m_{\sigma })\sigma ({\bf r}) &=&-g_{\sigma }\rho _{s}({\bf r}),
\label{kleingordons} \\
(-\Delta +m_{\omega })\omega ({\bf r}) &=&g_{\omega }\rho ({\bf r}),
\label{kleingordono} \\
(-\Delta +m_{\rho })\rho _{3}({\bf r}) &=&g_{\rho }(\rho _{n}({\bf r)}-\rho
_{p}({\bf r)}),  
\label{kleingordonr} \\
-\Delta A_{0}({\bf r}) &=&e^{2}\rho _{c}({\bf r),}  
\label{laplace}
\end{eqnarray}
$m_{\sigma }$, $m_{\omega}$, and $m_{\rho }$ are the masses of the 
$\sigma $-meson, the $\omega$-meson, and the $\rho $-meson, respectively. 
$g_{\sigma }$, $g_{\omega }$, and $g_{\rho }$ are the corresponding
coupling constants for the mesons to the nucleon, and $e^{2}/4\pi =1/137.036$.
$\rho _{c}({\bf r)}$ is the charge density. The interactions that result from
the Yukawa and Coulomb type Green's function are
\begin{eqnarray}
v_{s}({\bf r,r}^{\prime }) &=&-\frac{g_{\sigma }^{2}}{4\pi }\frac{%
e^{-m_{\sigma }|{\bf r-r}^{\prime }|}}{|{\bf r-r}^{\prime }|},
\label{yukawas} \\
v_{v}({\bf r,r}^{\prime }) &=&\frac{g_{\omega }^{2}}{4\pi }\frac{%
e^{-m_{\omega }|{\bf r-r}^{\prime }|}}{|{\bf r-r}^{\prime }|}+{\bf \vec{\tau}%
\,\vec{\tau}}^{\prime }\frac{g_{\rho }^{2}}{4\pi }\frac{e^{-m_{\rho }|{\bf %
r-r}^{\prime }|}}{|{\bf r-r}^{\prime }|}+\frac{e^{2}}{4\pi }\frac{1}{|{\bf %
r-r}^{\prime }|},  
\label{yukawav}
\end{eqnarray}
where ${\bf \vec{\tau}}$ are the isospin matrices.

In the mean-field approximation the mesons are introduced only 
on a phenomenological level as carriers of the fields. On a more
fundamental level, one should certainly try to relate these effective
mesons to physical mesons. For the
vector mesons $\omega $ and $\rho $ this relation is relatively well
established. The effective masses computed from the experimental 
properties of finite nuclei are very close to the free values. In fact,
in many successful parameter sets reported in the literature, one 
finds that the bare values have been used for the masses of the 
$\omega $ and $\rho $ mesons. This choice reduces the number of free
model parameters by 2. In the scalar sector, however, the
phenomenological level of the model becomes evident. A physical
scalar isoscalar meson has not been observed. On the other hand, it is well
known that the one-pion exchange determines the nucleon-nucleon
interaction at large distances. In the region of intermediate distances 
the interaction is predominantly the correlated two-pion exchange
and the two-pion exchange with a $\Delta $-particle in the intermediate 
state. Both processes result in a parity-conserving mean field, 
which in the RMF model is described by the
scalar field $S({\bf r})$. In principle, the Lorentz structure would allow a
pseudo-scalar field which originates from the one-pion exchange. On the 
Hartree level, however, this contribution vanishes due to parity 
conservation, and therefore the pion contributes  
only through the two-pion exchange.

Summarizing in a formal way our considerations,  
RMF can be formulated as an effective
field theory, starting from the Lagrangian density
originally introduced by Walecka  \cite{SW.86}:
\begin{eqnarray}
{\cal L} &=&\bar{\psi}\left( i\gamma \cdot \partial -m\right) \psi ~+~\frac{1%
}{2}(\partial \sigma )^{2}-\frac{1}{2}m_{\sigma }\sigma ^{2}  \nonumber \\
&&-~\frac{1}{4}\Omega _{\mu \nu }\Omega ^{\mu \nu }+\frac{1}{2}m_{\omega
}^{2}\omega ^{2}~-~\frac{1}{4}{\vec{{\rm R}}}_{\mu \nu }{\vec{{\rm R}}}^{\mu
\nu }+\frac{1}{2}m_{\rho }^{2}\vec{\rho}^{\,2}~-~\frac{1}{4}{\rm F}_{\mu \nu
}{\rm F}^{\mu \nu }  \nonumber \\
&&-~g_{\sigma }\bar{\psi}\sigma \psi ~-~g_{\omega }\bar{\psi}\gamma \cdot
\omega \psi ~-~g_{\rho }\bar{\psi}\gamma \cdot \vec{\rho}\vec{\tau}\psi ~-~e%
\bar{\psi}\gamma \cdot A\frac{(1-\tau _{3})}{2}\psi \;.  
\label{lagrangian}
\end{eqnarray}
Vectors in isospin space are denoted by arrows, and bold-faced
symbols will indicate vectors in ordinary three-dimensional space. 
$\Omega ^{\mu \nu }$, $\vec{R}^{\mu \nu }$, and $F^{\mu \nu }$ are the field
tensors of the vector fields $\omega $, $\rho $, and of the photon: 
\begin{eqnarray}
\Omega ^{\mu \nu } &=&\partial ^{\mu }\omega ^{\nu }-\partial ^{\nu }\omega
^{\mu } \\
\vec{R}^{\mu \nu } &=&\partial ^{\mu }\vec{\rho}^{\,\nu }-\partial ^{\nu }%
\vec{\rho}^{\,\mu } \\
F^{\mu \nu } &=&\partial ^{\mu }A^{\nu }-\partial ^{\nu }A^{\mu }.
\end{eqnarray}
If the bare masses $m$, $m_{\omega }$, and $m_{\rho }$ are used for the
nucleons and the $\omega $ and $\rho $ mesons, there are only four free
model parameters: $m_{\sigma }$, $g_{\sigma }$, $g_{\omega }$ 
and $g_{\rho }$. Their values can be adjusted to the experimental 
data on just few spherical nuclei. This simple model, however, is
not flexible enough for a quantitative description of properties of 
complex nuclear systems. An effective density
dependence is introduced \cite{BB.77} by replacing the quadratic 
$\sigma $-potential $\frac{1}{2}m^2_{\sigma }\sigma ^{2}$ 
with a quartic potential $U(\sigma )$ 
\begin{equation}
U(\sigma )~=~\frac{1}{2}m_{\sigma }^{2}\sigma ^{2}+\frac{1}{3}g_{2}\sigma
^{3}+\frac{1}{4}g_{3}\sigma ^{4}.  \label{usigma}
\end{equation}
The potential includes the nonlinear $\sigma $
self-interaction, with two additional  parameters $g_{2}$ and $g_{3}$. 
The corresponding Klein-Gordon equation (\ref{kleingordons})
becomes nonlinear, with a $\sigma $%
-dependent mass $m^2_{\sigma }(\sigma )=$ $m^2_{\sigma }+g_{2}\sigma
+g_{3}\sigma ^{2}$. More details on the relativistic mean-field
formalism can be found in Refs.~\cite{Rin.96,SW.86,Rei.89,Ser.92}.

The essential structure of the model is based on relativity. In a description
of the ground-state properties of nuclei we are certainly faced
with the question: why is it necessary to use a relativistic formulation? In
fact, the kinetic energies and the Fermi momenta are relatively small 
compared to the rest mass of the nucleon. Therefore, relativistic
kinematics can be neglected. The Dirac equation, however, contains more than
just the difference $V({\bf r})-S({\bf r})$. 
Compared to an equivalent Schr\"{o}dinger equation with a potential 
well of $\sim $50 MeV 
(small compared to the nucleon rest mass of 939 MeV), the Dirac
equation contains two potentials: $V({\bf r})$ and $-S({\bf r})$, which are
both large ($\sim 350$ and $\sim -400$ MeV) and opposite in sign. Therefore,
one needs a relativistic framework in order to properly describe the  
dynamics of these strong potentials. In the equation (\ref{dirac2})
for the large component $f$, only the difference between 
of $V$ and $S$ enters, and this potential is, in fact, small compared to the
nucleon rest mass. In the lower equation for the small components $g$,
the sum  $V+S$ occurs. This term is large and it produces the strong
spin-orbit potential, which is essential in nuclear structure calculations.
One of the advantages of working in a relativistic framework is also being
able to determine the strength and the shape of the 
spin-orbit term self-consistently.
This is especially important for a correct description of spin-orbit 
splittings in regions of nuclei far from the valley of $\beta$ stability,
where the extrapolation of effective strength parameters becomes 
questionable. Another example of the importance of describing the 
single-nucleon wave functions with Dirac spinors, is the occurrence of 
approximate pseudo-spin symmetry in nuclear spectra. The symmetry 
has its origin in the approximate cancellation of the scalar 
and vector potentials in the single-nucleon Dirac equation\cite{Gin.97}.

It should also be emphasized that the smallness of $V-S$ leads to
relatively small Fermi momenta and allows, in principle, a non-relativistic
reduction of the Dirac equation to a Schr\"{o}dinger equation with
momentum-dependent potentials. The resulting non-relativistic theory with
additional spin- and momentum-dependent terms requires, in general,
more adjustable parameters, and one can argue that 
its predictive power is reduced as compared to a fully 
relativistic theory.

\subsection{Relativistic Theory with Pairing Correlations}

The inclusion of pairing correlations is essential for a correct
description of structure phenomena in spherical open-shell nuclei 
and in deformed nuclei.  The relativistic mean field theory, as
described in the previous section, does not include such correlations.
The Dirac Hamiltonian contains only single-particle field operators
with the structure $\psi ^{\dagger }\psi $; pairing correlations can 
only be described in a generalized single-particle theory by 
field operators $\psi ^{\dagger}\psi ^{\dagger }$ and $\psi \psi $,
which do not conserve the particle number.
Pairing correlations are therefore often included in a phenomenological way 
with the simple BCS approximation \cite{GRT.90}. However, the BCS model 
presents only a poor approximation for
exotic nuclei far from the valley of $\beta$-stability. The physics of
drip-line nuclei necessitates a unified and self-consistent treatment of
mean-field and pairing correlations. In the non-relativistic approach,
properties of drip line nuclei are consistently described in the framework
of Hartree-Fock-Bogoliubov (HFB) theory \cite{DNW.96}. The relativistic
extension of the HFB theory was introduced in Ref. \cite{KR.91}. The 
starting point is again the Lagrangian (\ref{lagrangian}). By performing
the quantization of the meson fields, the
relativistic Hartree-Bogoliubov equations are
derived using Green's function techniques.

As in non-relativistic HFB theory, the ground state of a nucleus 
$|\Phi >$ is described as the vacuum with respect to independent 
quasi-particle operators, which are defined by a unitary Bogoliubov 
transformation of the single-nucleon creation and annihilation operators. 
The generalized single-nucleon Hamiltonian contains two average potentials: 
the self-consistent Hartree-Fock field $\hat{\Gamma}$ which encloses all the 
long range {\it ph} correlations, and a pairing
field $\hat{\Delta}$ which sums up the {\it pp}-correlations. 
In the relativistic extension the Hartree approximation is
employed for the self-consistent mean-field, and the resulting Relativistic
Hartree-Bogoliubov (RHB) equations read 
\begin{equation}
\left( 
\begin{array}{cc}
\hat{h}_{D}-m-\lambda  & \hat{\Delta} \\ 
-\hat{\Delta}^{\ast } & -\hat{h}_{D}+m+\lambda 
\end{array}
\right) \left( 
\begin{array}{c}
U_{k}({\bf r}) \\ 
V_{k}({\bf r})
\end{array}
\right) =E_{k}\left( 
\begin{array}{c}
U_{k}({\bf r}) \\ 
V_{k}({\bf r})
\end{array}
\right)   
\label{HFB}
\end{equation}
where $\hat{h}_{D}$ is the single-nucleon Dirac Hamiltonian (\ref{dirac1}),
and $m$ is the nucleon mass. The chemical potential $\lambda $ has to be
determined by the particle number subsidiary condition, in order that the
expectation value of the particle number operator in the ground state equals
the number of nucleons. The column vectors denote the quasi-particle wave
functions, and $E_{k}$ are the quasi-particle energies.

As it is well known, from the non-relativistic Hartree-Fock-Bogoliubov 
theory and the detailed discussion in Ch. 7 of Ref. \cite{RS.80}, 
for each of the positive quasiparticle energies $E_k>0$ with eigenvector 
$(U^{}_k,V^{}_k)$ exists an eigensolution $-E_k$ with eigenvector 
$(V^*_k,U^*_k)$. 
One has to choose for each quantum number $k$ one of these solutions.
In Hartree-Fock calculations this corresponds to the choice of a 
certain occupation-pattern for the Slater determinant. For the calculation
of the ground state of an even-even nucleus one occupies the lowest
levels starting from the bottom of the well, i.e. one chooses the lowest 
A single-particle levels. In the HFB case the ground-state 
of an even-even system is obtained by choosing only the positive 
quasi-particle energies (see Ref. \cite{RS.80} for exceptions of this rule in
the case of rotating nuclei). To describe odd nuclei one has to block 
at least one level $k_0$. This means that one has to choose for $k_0$ 
the negative eigenvalue $-E_{k_0}$ with the eigenvector 
$(V^*_{k_0},U^*_{k_0})$. Therefore, in the 
following equations summations over $k$ run only over half of the 
eigensolutions of the  RHB-equation (\ref{HFB}) chosen according 
to this prescription.

The RHB equations are solved self-consistently, with potentials
determined in the mean-field approximation from solutions of Klein-Gordon
equations 
\begin{eqnarray}
\left[ -\Delta +m_{\sigma }^{2}\right] \,\sigma ({\bf r}) &=&-g_{\sigma
}\,\rho _{s}({\bf r})-g_{2}\,\sigma ^{2}({\bf r})-g_{3}\,\sigma ^{3}({\bf r})
\label{messig} \\
\left[ -\Delta +m_{\omega }^{2}\right] \,\omega ^{0}({\bf r}) &=&g_{\omega
}\,\rho _{v}({\bf r})  \label{mesome} \\
\left[ -\Delta +m_{\rho }^{2}\right] \,\rho ^{0}({\bf r}) &=&g_{\rho }\,\rho
_{3}({\bf r})  \label{mesrho} \\
-\Delta \,A^{0}({\bf r}) &=&e\,\rho _{p}({\bf r}),  \label{photon}
\end{eqnarray}
for the sigma meson, omega meson, rho meson and photon field, respectively.
Due to charge conservation, only the 3rd-component of the isovector rho
meson contributes. The source terms in equations (\ref{messig}) to (\ref
{photon}) are sums of bilinear products of baryon amplitudes 
\begin{eqnarray}
\rho _{s}({\bf r}) &=&\sum\limits_{k}V_{k}^{\dagger }({\bf r})\gamma
^{0}V_{k}({\bf r}),  \label{equ.2.3.h} \\
\rho _{v}({\bf r}) &=&\sum\limits_{k}V_{k}^{\dagger }({\bf r})V_{k}(%
{\bf r}), \\
\rho _{3}({\bf r}) &=&\sum\limits_{k}V_{k}^{\dagger }({\bf r})\tau
_{3}V_{k}({\bf r}), \\
\rho _{{\rm em}}({\bf r}) &=&\sum\limits_{k}V_{k}^{\dagger }({\bf r}){%
\frac{{1-\tau _{3}}}{2}}V_{k}({\bf r}),
\end{eqnarray}
where the sums run over all the all the quasiparticle states with
positive energy in even-even nuclei. In odd systems at least one
level $k_0$ has to be blocked, i.e. the corresponding eigenvector 
$V^{}_{k_0}(r)$ has to be replaced by $U^{*}_{k_0}(r)$.

The pairing field $\hat{\Delta}$ in (\ref{HFB}) is an integral operator with
the kernel 
\begin{equation}
\Delta _{ab}({\bf r},{\bf r}^{\prime })={\frac{1}{2}}\sum%
\limits_{c,d}V_{abcd}({\bf r},{\bf r}^{\prime }){\bf \kappa }_{cd}({\bf r},%
{\bf r}^{\prime }),  
\label{equ.2.5}
\end{equation}
where $a,b,c,d$ denote quantum numbers which specify the Dirac indices of the
spinors, $V_{abcd}({\bf r},{\bf r}^{\prime })$ are matrix elements of a
general two-body interaction. 

The eigensolutions of Eq. (\ref{HFB}) form a set of orthonormal
single quasi-particle states. The corresponding eigenvalues are
the single quasi-particle energies. The self-consistent iteration procedure
is performed in the basis of quasi-particle states. As shown by 
Bloch and Messiah in \cite{BM.62}, any Hartree-Fock-Bogoliubov wave 
function can also be represented in the form of a BCS function. For
systems with an even number of particles
\begin{equation}
|\Phi \rangle~=~\prod_{\mu }
(u_{\mu }+v_{\mu }a_{\mu }^{\dagger }a_{\overline{\mu }}^{\dagger })|-\rangle,  \label{BCS}
\end{equation}
and for odd particle number
\begin{equation}
|\Phi_{\mu_0}\rangle~=~a^\dagger_{\mu_0}\prod_{\mu\neq\mu_0}
(u_\mu+v_\mu a_\mu^\dagger a_{\overline{\mu}}^{\dagger })|-\rangle,  
\label{BCS1}
\end{equation}
where $\mu_0$ is the level occupied by the odd particle (blocked level).

The operators $a_{\mu }^{\dagger }$ and $a_{\overline{\mu }}^{\dagger
}$ create particles in the canonical basis, and the occupation
probabilities are given by 
\begin{equation}
v_{\mu }^{2}=\frac{1}{2}\left( 1-\frac{\varepsilon _{\mu }^{{}}-m-\lambda }{%
\sqrt{(\varepsilon _{\mu }^{{}}-m-\lambda )^{2}+\Delta _{\mu }^{2}}}\right).
\label{occup}
\end{equation}
$\varepsilon _{\mu }^{{}}=\left\langle \mu |h_{D}|\mu \right\rangle $
and $\Delta _{\mu }^{{}}=$ $\left\langle \mu |\hat{\Delta}|\overline{\mu }%
\right\rangle $ are the diagonal elements of the Dirac single-particle
Hamiltonian and the pairing field in the canonical basis, respectively. 
In contrast to the BCS framework, however, neither of
this fields is diagonal in the canonical basis. The basis itself is
specified by the requirement that it diagonalizes 
the single particle density matrix 
$\hat{\rho}({\bf r,r}^{\prime })=
\sum\limits_{k}V^{}_{k}({\bf r})V^\dagger_{k}({\bf r}^{\prime })$. 
The transformation to the canonical basis determines the
energies and occupation probabilities of single-particle states, which
correspond to the self-consistent solution for the ground state of a nucleus.

According to the Bloch-Messiah theorem the canonical basis is connected
to the quasiparticle basis through a unitary transformation (usually
called $C$-transformation). Only in the HF+BCS case this
transformation is proportional to unity. Therefore, the blocked 
level $\mu_0$ in Eq. (\ref{BCS1}), which is a basis vector of
the canonical basis, is in general different from the blocked 
quasiparticle level $k_0$ discussed after Eq. (\ref{HFB}). 
However, according to the Bloch-Messiah theorem a density matrix 
$\hat{\rho}({\bf r,r}^{\prime })$, obtained after blocking the 
quasiparticle level $k_0$, has always one eigenvalue 1 and one 
eigenvalue 0 in addition to the doubly degenerate eigenvalues
$v^2_\mu$. Neglecting degeneracies, there is a unique relation
between the quasiparticle level $k_0$ and the blocked level
$\mu_0$ in the canonical basis. In case of symmetries both
levels $\mu_0$ and $k_0$ have identical quantum numbers while in 
cases where the matrix $C$ is close to unity, the corresponding 
wavefunctions are similar. 

In many applications of the relativistic mean-field model pairing
correlations have been taken into account in a very phenomenological way in
the BCS model with the monopole pairing force, adjusted to the experimental
odd-even mass differences. This framework obviously cannot be applied to the
description of the coupling to the particle continuum in nuclei close to
drip-line. The question therefore arises, which pairing interaction 
$V_{abcd}({\bf r},{\bf r^{\prime }})$ should be used in Eq. (18). In
Ref. \cite{KR.91} a fully relativistic derivation of the
pairing force has been developed, starting from the Lagrangian (1). Using
the Gorkov factorization technique, it has been possible to demonstrate that
the pairing interaction results from the one-meson exchange ($\sigma $-, $%
\omega $- and $\rho $-mesons), and is therefore of the form
(\ref{yukawas}) and (\ref{yukawav}). In practice, however, 
it turns out that the
pairing correlations calculated in this way, with coupling constants taken
from the standard parameter sets of the RMF model, are too strong. The
repulsion produced by the exchange of vector mesons at short distances
results in a pairing gap at the Fermi surface that is by a factor three too
large. However, as has been argued in many applications of the
Hartree-Fock-Bogoliubov theory, there is no real reason to use the same
effective forces in both the particle-hole and particle-particle channels.
In the ladder approximation, the effective interaction contained in the
mean-field $\hat{\Gamma}$ is a G-matrix, the sum over all ladder diagrams.
The effective force in the {\it pp} channel, i.e. in the pairing potential 
$\hat{\Delta}$, should be the K matrix, the sum of all diagrams irreducible
in the {\it pp}-direction (see for instance \cite{Mig.67}).
Since very little is known about this matrix in the relativistic approach, 
in most applications of the RHB model a phenomenological pairing 
interaction has been used, the pairing part of the Gogny force, 
\begin{equation}
V^{pp}(1,2)~=~\sum_{i=1,2}e^{-(({\bf r}_{1}-{\bf r}_{2})/{\mu _{i}}%
)^{2}}\,(W_{i}~+~B_{i}P^{\sigma }-H_{i}P^{\tau }-M_{i}P^{\sigma }P^{\tau }),
\end{equation}
with the set D1S \cite{BGG.84} for the parameters $\mu _{i}$, $W_{i}$, 
$B_{i}$, $H_{i}$ and $M_{i}$ $(i=1,2)$.

Although this procedure formally breaks the Lorentz structure of 
the equations, one has to keep in mind that pairing in itself is a
completely non-relativistic phenomenon. The essential 
point in a relativistic description of nuclei 
is the difference between the scalar and the vector density, 
which yields saturation, together with the approximate cancellation 
of the scalar attraction by the vector repulsion, 
for the large components in the Dirac equation. The addition 
of the scalar and vector potentials for the small components,
results in large energy splitting between spin-orbit partners. 
The admixture of small components through the kinetic 
term ${\bf\sigma p}$ in the off-diagonal part of the Dirac equation, 
has no counterpart in the pairing field. Relativistic 
effects are only important for the mean field part of Hartree-
Bogoliubov theory. The pairing density $\kappa$ and the corresponding 
pairing field, result from the scattering of pairs in the
vicinity of the Fermi surface. The pairing density is concentrated  
in an energy window of less than 50 MeV around the 
Fermi level, i.e. the contributions from the small components to 
the pairing tensor $\kappa$ are extremely small. The word {\it 
relativistic} in RHB, therefore, applies always the the Hartree 
{\it ph}-channel of this theory; the Bogoliubov {\it pp}-channel 
is never relativistic. Consequently, it is justified to approximate 
the pairing force by the best currently available 
non-relativistic interaction: the pairing part of the Gogny-force. 
This is certainly more realistic than a one-meson exchange 
force in the pairing channel, since this type of interactions 
have never been optimized for the description
of pairing properties. 

\subsection{Deformed Dirac-Hartree-Bogoliubov Equations}

The self-consistent solution of the Dirac-Hartree-Bogoliubov
integro-differential eigenvalue equations and Klein-Gordon equations for the
meson fields determines the nuclear ground state. For systems with spherical
symmetry, i.e. single closed-shell nuclei, the coupled system of equations
has been solved using finite element methods in coordinate space 
\cite{PVL.97,LVP.98,LVR.97,PVR.97,VPLR.98}, and by expansion in a basis of
spherical harmonic oscillator~\cite{LVR.98,VLR.98,GEL.96}. For deformed
nuclei the present version of the model does not include solutions in
coordinate space. The Dirac-Hartree-Bogoliubov equations and the equations
for the meson fields are solved by expanding the nucleon spinors 
$U_k({\bf r})$ and $V_k({\bf r})$, 
and the meson fields, in terms of the eigenfunctions
of a deformed axially symmetric oscillator potential~\cite{GRT.90}. 
For nuclei at the drip-lines, however, solutions in configurational
representation might not provide an accurate description of properties that
crucially depend on the spatial extension of nucleon densities, as for
example nuclear radii. In less exotic nuclei on the
neutron rich side, or for proton-rich nuclei, an expansion in a large
oscillator basis should provide sufficiently accurate solutions. In
particular, proton-rich nuclei are stabilized by the Coulomb barrier which
tends to localize the proton density in the nuclear interior and thus
prevents the formation of objects with extreme spatial extension.

The current version of the model describes axially symmetric deformed
shapes. For this geometry the rotational symmetry is
broken and the total angular momentum $j$ is no longer a good
quantum number. However, the densities and fields are still invariant with
respect to a rotation around the symmetry axis, which is taken to be the
z-axis. It is then natural to use cylindrical coordinates to
represent the solutions of the RHB equations. In particular, the
single-nucleon spinors $U({\bf r})$ and $V({\bf r})$ in (9) are
characterized by the quantum numbers [$\Omega ,\pi ,t_{3}$]. $\Omega $ is
the eigenvalue of the symmetry operator $j_{z}$ (the projection of the total
angular momentum on the symmetry axis), $\pi $ is the parity, and $t_{3}$
the isospin projection. The two Dirac spinors $U({\bf r})$ and $V({\bf r})$
are defined as 
\begin{equation}
U({\bf r},s,t)\ =\ 
\left( \begin{array}{c}
f_{U}({\bf r},s,t) \\ ig_{U}({\bf r},s,t)  
\end{array} \right)\ = \
{\frac{1}{\sqrt{2\pi }}}\left( 
\begin{array}{c}
f_{U}^{+}(z,r_{{\bot }})\ e^{i(\Omega -1/2)\varphi } \\ 
f_{U}^{-}(z,r_{{\bot }})\ e^{i(\Omega +1/2)\varphi } \\ 
ig_{U}^{+}(z,r_{{\bot }})\,e^{i(\Omega -1/2)\varphi } \\ 
ig_{U}^{-}(z,r_{{\bot }})\,e^{i(\Omega +1/2)\varphi }
\end{array}
\right) \ \chi _{t_{3}}(t)
\label{U}
\end{equation}
and 
\begin{equation}
V({\bf r},s,t)\ =\ 
\left( \begin{array}{c}
f_{V}({\bf r},s,t) \\ ig_{V}({\bf r},s,t)  
\end{array} \right)\ = \
{\frac{1}{\sqrt{2\pi }}}\left( 
\begin{array}{c}
f_{V}^{+}(z,r_{{\bot }})\ e^{i(\Omega -1/2)\varphi } \\ 
f_{V}^{-}(z,r_{{\bot }})\ e^{i(\Omega +1/2)\varphi } \\ 
ig_{V}^{+}(z,r_{{\bot }})\,e^{i(\Omega -1/2)\varphi } \\ 
ig_{V}^{-}(z,r_{{\bot }})\,e^{i(\Omega +1/2)\varphi }
\end{array}
\right) \ \chi _{t_{3}}(t)
\label{V}
\end{equation}
\bigskip

The basis expansion method starts from the deformed axially symmetric
oscillator potential
\begin{equation}
V_{osc}(z,r_{{\bot }})\quad =\quad {\frac{{1}}{{2}}}M\omega _{z}^{2}z^{2}\
+\ {\frac{{1}}{{2}}}M\omega _{{\bot }}^{2}r_{{\bot }}^{2}.
\end{equation}
Imposing volume conservation, the two oscillator frequencies $\hbar \omega _{%
{\bot }}$ and $\hbar \omega _{z}$ can be expressed in terms of a deformation
parameter $\beta _{0}$: 
\begin{eqnarray}
\hbar \omega _{z}\ &=&\ \hbar \omega _{0}\ \exp (-\sqrt{\frac{5}{{4\pi }}}%
\beta _{0}) \\
\hbar \omega _{{\bot }}\ &=&\ \hbar \omega _{0}\ \exp (+{{\frac{1}{2}}\sqrt{%
\frac{5}{{4\pi }}}}\beta _{0}).
\end{eqnarray}
The corresponding oscillator length parameters are 
\begin{equation}
b_{z}\ =\ \sqrt{\frac{\hbar }{{M\omega _{z}}}}\quad {\rm and}\quad b_{{\bot }%
}\ =\ \sqrt{\frac{\hbar }{{M\omega _{{\bot }}}}}
\end{equation}
Because of volume conservation, $b_{{\bot }}^{2}b_{z}=b_{0}^{3}$.
The basis is now specified by the two constants $\hbar \omega _{0}$ and $%
\beta _{0}$. The eigenfunctions of the deformed harmonic oscillator
potential are characterized by the quantum numbers 
\begin{equation}
\mid \alpha >\ =\ \mid n_{z},n_{r},m_{l},m_{s}>,
\label{quantum}
\end{equation}
where $m_{l}$ and $m_{s}$ are the components of the orbital angular momentum
and the spin along the symmetry axis, respectively. 
The eigenvalue of $j_{z}$, which is
a conserved quantity in these calculations, is 
\begin{equation}
\Omega \ =\ m_{l}+m_{s},
\end{equation}
and the parity is given by  
\begin{equation}
\pi \ =\ (-)^{n_{z}+m_{l}} .
\end{equation}
The eigenvectors can be explicitly written as a product of functions  
\begin{eqnarray}
\Phi _{\alpha }(z,r_{{\bot }},\varphi ,s,t)\ &=&\ \phi _{n_{z}}(z)\ \phi
_{n_{r}}^{m_{l}}(r_{{\bot }}){\frac{{1}}{\sqrt{2\pi }}}e^{im_{l}\varphi
}\chi _{ms}(s)\chi _{t_{\alpha }}(t) \\
\ &=&\Phi _{\alpha }({\bf r},s)\,\chi _{t_{\alpha }}(t),
\end{eqnarray}
with 
\begin{eqnarray}
\phi _{n_{z}}(z)\ \ &=&\ {\frac{{N_{n_{z}}}}{\sqrt{b_{z}}}}\ H_{n_{z}}(\zeta
)\,e^{-\zeta ^{2}/2} \\
\phi _{n_{r}}^{m_{l}}(r_{{\bot }})\ &=&\ {\frac{{N_{n_{r}}^{m_{l}}}}{{b_{{%
\bot }}}}}\ \sqrt{2}\,\eta ^{m_{l}/2}\,L_{n_{r}}^{m_{l}}(\eta )\,e^{-\eta /2},
\end{eqnarray}
and 
\begin{equation}
\zeta \ =\ z/b_{z},\qquad \eta \ =\ r_{{\bot }}^{2}/b_{{\bot }}^{2}.
\end{equation}
$H_{n}(\zeta )$ and $L_{n}^{m}(\eta )$ denote the Hermite
and associated Laguerre polynomials, respectively. The
normalization constants are given by 
\begin{equation}
N_{n_{z}}\ =\ {\frac{1}{\sqrt{\sqrt{\pi }2^{n_{z}}n_{z}!}}}\quad {\rm and}%
\quad N_{n_{r}}^{m_{l}}\ =\ \sqrt{\frac{{n_{r}!}}{{(n_{r}+m_{l})!}}}.
\end{equation}
For the nucleon spinors $U({\bf r})$ and $V({\bf r})$ in Eqs. (\ref{U})
and (\ref{V}), the following expansion in terms of the eigenfunctions is used
\begin{eqnarray}
f_{U(V)}({\bf r},s,t)&=& 
\sum_\alpha^{\alpha_{max}} f_{U(V)\alpha}\,
\Phi_{\alpha}({\bf r},s)\,\chi _{t_{i}}(t) 
\label{expansion}\\
g_{U(V)}({\bf r},s,t)&=&
\sum_{\tilde{\alpha}}^{{\tilde{\alpha}}_{max}}g_{U(V)\tilde{\alpha}}\,
\Phi _{\tilde{\alpha}}({\bf r},s)\,\chi _{t_{i}}(t).
\end{eqnarray}
In order to avoid the occurrence of spurious states, the quantum
numbers $\alpha _{max}$ and ${\tilde{\alpha}}_{max}$ are chosen in such a
way, that the corresponding major quantum numbers $N=n_{z}+2n_{\rho }+m_{l}$
are not larger than $N_{F}+1$ for the expansion of the small components, and
not larger than $N_{F}$ for the expansion of the large components \cite{GRT.90}. \medskip 

The solutions for the fields of massive mesons are also obtained with an
expansion in a deformed oscillator basis. It is convenient to use the same
deformation parameter $\beta _{0}$ as for the fermion fields, and to choose
the oscillator length $b_{B}=b_{0}/\sqrt{2}$ : 
\begin{equation}
\phi (z,r_{{\bot }})\ =\ {\frac{1}{{b_{B}^{3/2}}}}e^{-\zeta ^{2}/2-\eta
/2}\sum_{n_{z}n_{r}}^{N_{B}}\phi _{n_{z}n_{r}}N_{n_{z}}\,H_{n_{z}}(\zeta )\ 
\sqrt{2}\,L_{n_{r}}^{0}(\eta ),
\end{equation}
with $\zeta =\sqrt{2}z/b_{z}$ and $\eta =2r_{{\bot }}^{2}/b_{{\bot }}^{2}$.
Inserting this ansatz into the Klein-Gordon equation, an
inhomogeneous set of linear equations is obtained
\begin{equation}
\sum_{{n_{z}}^{\prime }{n_{r}}^{\prime }}^{N_{B}}{\cal H}_{n_{z}n_{r}{n_{z}}%
^{\prime }{n_{r}}^{\prime }}\ \phi _{{n_{z}}^{\prime }{n_{r}}^{\prime }}\ =\
s_{n_{z}n_{r}}^{\phi },
\end{equation}
with the matrix elements 
\begin{eqnarray}
{\cal H}_{n_{z}n_{r}{n_{z}}^{\prime }{n_{r}}^{\prime }}&=&
m_\phi^2 \delta_{n_r{n_r}^\prime}\delta_{n_z{n_z}^\prime} 
\\
&+&\frac{1}{2b_z^2}
\left((2n_z+1)\ \delta_{n_z{n_z}^\prime}\ -\  
{n_z}^\prime\delta_{{n_z}^\prime n_z+1}\ -\ 
n_z\ \delta_{n_z{n_z}^\prime+1} \right)
\delta_{n_r{n_r}^\prime}
\nonumber\\
&+&\frac{1}{~b_\bot^2}
\left( (2n_r+1)\,\delta_{n_r{n_r}^\prime}\ +\  
{n_r}^\prime\delta_{{n_r}^\prime n_r+1}\ +\ 
n_r\ \delta_{n_r{n_r}^\prime+1} \right)
\delta_{n_z{n_z}^\prime}
\nonumber
\end{eqnarray}

Most of our calculations are performed by expansion in 12 oscillator shells
for the fermion fields, and 20 shells for the boson fields. In order to test
the convergence, critical cases are also calculated with 14 shells for the
fermion fields.

\subsection{Systems with Odd Number of Particles}

In the study of proton emitters nuclei with both even and odd number 
of protons (neutrons) must be calculated. The RHB wave function 
of a system with an odd number of particles is represented 
by the product state (\ref{BCS1}) in the canonical basis. 
Therefore, for a nucleus with odd proton (neutron) number, one 
proton (neutron) level must be blocked. If both Z and N
are odd, one proton level and one neutron level must 
be blocked. As described in Ref. \cite{RS.80}, a quasiparticle level
$k$ is blocked by exchanging the two eigensolutions
$(U_k^{},V_k^{})$ and $(V_k^*,U_k^*)$ of the RHB-equations (\ref{HFB}). 
The blocking procedure is performed in each step of the 
self-consistent iteration. The resulting density matrix and 
pairing tensor, in the oscillator basis, read \cite{RBM.70,EMR.80}
\begin{eqnarray}
\rho_{\alpha\alpha^\prime}^{(k)}&=&\rho_{\alpha\alpha^\prime}^{}
~+~U_{\alpha k}^{}U_{\alpha^\prime k}^{*}
~-~V_{\alpha k}^{*}V_{\alpha^\prime k}^{},
\\
\kappa_{\alpha\alpha^\prime}^{(k)}&=&\kappa_{\alpha\alpha^\prime}^{}
~+~U_{\alpha k}^{}V_{\alpha^\prime k}^{*}
~-~V_{\alpha k}^{*}U_{\alpha^\prime k}^{},
\end{eqnarray}
where 
\begin{equation}
\rho_{\alpha\alpha^\prime}^{}~=~(V^* V^T)_{\alpha\alpha^\prime}^{}
~~~~~~{\rm and}~~~~~~
\kappa_{\alpha\alpha^\prime}^{}~=~(V^* U^T)_{\alpha\alpha^\prime}^{}
\end{equation}
are the corresponding matrices for the system without blocking.
Each blocked level, of course, produces a different solution for 
the ground state. It might happen that several quasiparticle
levels available to the odd proton (neutron) are found close 
in energy. This case is especially important for deformed nuclei.
In order to determine the self-consistent solution for
the ground state, each available level has to be blocked in turn.
The ground state then corresponds to the self-consistent
solution with lowest energy. Separation
energies and spectroscopic factors of proton emitters are 
calculated from the self-consistent ground state RHB solutions
for the parent system (odd Z) and daughter nucleus (even Z).
Of course, both nuclei can have an odd number of neutrons, 
in which case calculations with blocked neutron levels must 
be performed. 

\bigskip
%

\section{Proton drip-line and deformed ground-state proton emitters}


In the present work the relativistic Hartree-Bogoliubov theory 
is applied in the description of ground-state properties
of deformed proton-rich odd Z nuclei in
the region 53$\leq$Z$\leq$69. In particular, the RHB model is used to study
the location of the proton drip-line, the ground-state quadrupole
deformations and one-proton separation energies at and beyond the drip-line,
the deformed single-particle orbitals occupied by the odd valence proton,
and the corresponding spectroscopic factors.

The input parameters of the RHB model are the coupling constants and the
masses for the effective mean-field Lagrangian, and the effective
interaction in the pairing channel. As we have done in most of our recent
studies, we use the NL3 effective interaction \cite{LKR.97} for the RMF
Lagrangian. Properties calculated with NL3 indicate that this is probably
the best effective interaction so far, both for nuclei at and away from the
line of $\beta $-stability. A very recent systematic 
theoretical study of ground-state properties of 
more than 1300 even-even isotopes has shown excellent agreement 
with experimental data \cite{LRR.98}. For the pairing field we employ the
pairing part of the Gogny interaction with the parameter set D1S \cite
{BGG.84}. As we have already discussed in the previous section, 
this force has been very carefully adjusted to the pairing
properties of finite nuclei all over the periodic table.

In Figs. \ref{figA} and \ref{figB} the one-proton separation energies 
\begin{equation}
S_{p}(Z,N)=B(Z,N)-B(Z-1,N),  \label{sep}
\end{equation}
are displayed for the odd-Z nuclei $53\leq Z\leq 69$, as function of the
number of neutrons. The model predicts the drip-line nuclei: $^{110}$I, $%
^{115}$Cs, $^{118}$La, $^{124}$Pr, $^{129}$Pm, $^{134}$Eu, $^{139}$Tb, $%
^{146}$Ho, and $^{152}$Tm. For nuclei with $Z>50$, the superposition of the
Coulomb and centrifugal potentials results in a relatively high potential
energy barrier, through which the tunneling of the odd proton proceeds. For
the ground-state proton emission to occur, the valence proton must penetrate
the wide potential barrier, and this process competes with $\beta ^{+}$
decay. The half-life of the decay strongly depends on the energy of the
proton, i.e. on the $Q_{p}$ value, and on the angular momentum of the proton
state. The $Q_{p}$ dependence has been nicely illustrated in Ref. \cite
{ASN.97}, where proton partial decay half-lives for $^{147}$Tm have been
calculated in the DWBA. It has been shown that, as the $Q_{p}$ value changes
from 0.5 MeV to 2.5 MeV, the half-life is reduced by more than 22 orders of
magnitude, from $10^{10}$ $sec$ to $10^{-12}$ $sec$ for the decay of the $%
1d_{3/2}$ state. This means that in many nuclei the ground-state proton
emission will not be observed immediately after the drip-line, because for
small values of $Q_{p}$ the total half-life will be completely dominated by $%
\beta ^{+}$ decay. Only for higher values of $Q_{p}$ is proton radioactivity
expected to dominate over $\beta ^{+}$ decay. By further decreasing the
number of neutrons, the odd proton energy increases rapidly and this results
in extremely short proton-emission half-lives. For a typical rare-earth
nucleus the $Q_{p}$ window in which ground-state proton decay can be
directly observed is about 0.8 -- 1.7 MeV \cite{ASN.97}. An interesting
point is that for nuclei at the drip-lines the precise values of the
separation energies should also depend on the isovector properties of the
spin-orbit interaction. It has been shown that the relativistic
mean-field model encloses a unique isovector dependence of the spin-orbit
term of the effective single-nucleon potential \cite{LVR.97}, very different
from the one included in the more traditional non-relativistic models based
on Skyrme ansatz. It would be particularly interesting to compare
predictions of various models based on the framework of the self-consistent
mean-field theory, as it was done, for example, in Ref. \cite{Naz.96} for
the structure of proton-drip nuclei around the doubly magic $^{48}$Ni.

In Figs. \ref{figA} and \ref{figB} we display the separation energies of
nuclei beyond the drip-line. The energy window extends to include those
nuclei for which a direct observation of ground-state proton emission is in
principle possible on the basis of calculated separation energies. For the
most probable proton emitters, in Table \ref{TabA} we enclose the
ground-state properties calculated in the RHB model. For each nucleus we
include the one-proton separation energy $S_{p}$, the quadrupole deformation 
$\beta _{2}$, the deformed single-particle orbital occupied by the odd
valence proton, and the corresponding theoretical spectroscopic factor. 
In order to determine the Nilsson quantum numbers and the spectroscopic
factors, we analyze the wavefunctions of the parent nucleus with
an odd proton number and that of the daughter nucleus with even proton
number in the canonical basis. Because of the different mean fields
and because of the blocking, the canonical basis-systems found in
this way are not completely identical for the even and for the odd case.
However, the wavefunctions and the occupation numbers $v^2_\mu$ are in 
most of the cases rather similar. The essential difference 
consists in the fact, that one of the pairs of the degenerate occupation 
numbers $v^2_{\mu_0}$ does not occur in the odd system. Instead of
that we find the occupation numbers 0 and 1. This level is obviously the 
blocked level $\mu_0$ in the canonical basis. The Nilsson quantum 
numbers of the odd-proton level are determined by the dominant component 
in the expansion of this wavefunction in terms of the anisotropic 
oscillator basis (\ref{quantum}). The spectroscopic factor
of the deformed odd-proton orbital is defined as the probability
that this state is found empty in the daughter nucleus with even number of 
protons. If one neglects the difference of the canonical basis for the 
even and the odd proton system, the spectroscopic factor of the level 
$\mu_0$ is calculated from the Eqs. (\ref{BCS}) and (\ref{BCS1}):
\begin{equation}
S_{\mu_0}~=~|\langle \Phi_{\mu_0} | a_{\mu_0}^\dagger | \Phi\rangle |^2
~=~u_{\mu_0}^2.
\end{equation}
where $u_{\mu_0}$ is determined in the canonical basis of the daughter
nucleus with even proton number.

The results of RHB calculations are compared with the
predictions of the finite-range droplet (FRDM) mass model:
the projection of the odd-proton angular
momentum on the symmetry axis and the parity of the odd-proton state 
$\Omega_{p}^{\pi }$ \cite{MNK.97}, the one-proton separation 
energy \cite{MNK.97},
and the ground-state quadrupole deformation \cite{MN.95}. The separation
energies are also compared with available experimental data on proton
transition energies. An excellent agreement is found between ground-state
quadrupole deformations calculated in the two theoretical models. In
particular, both models predict a transition from prolate to oblate shapes
in the Tm isotopes. On the other hand, the predictions for the deformed
orbitals occupied by the odd proton differ in many cases. For the I and Cs
isotopes, the RHB calculations predict the $3/2^{+}[422]$ and $1/2^{+}[420]$
proton orbitals, respectively. 
Exactly the opposite ordering of orbitals is calculated in the
macroscopic-microscopic model. The differences are also pronounced in 
$^{115,116}$La, $^{130,131}$Eu, and $^{145}$Tm. In order to determine which
is the last occupied proton orbital, the corresponding decay width for
proton emission should be calculated and compared with the experiment value. 
For spherical proton emitters this is a relatively simple task \cite{ASN.97},
but the extension of the formalism to deformed nuclei is rather involved.
Calculations based on the coupled-channel formalism for deformed proton
emitters have been reported for $^{109}$I and $^{113}$Cs \cite{BK.89}, and
more recently in Ref. \cite{MFL.98} for $^{113}$Cs. These calculations are
not based on a fully microscopic and self-consistent description of proton
unstable nuclei, but use a Woods-Saxon potential adjusted in such a way that
the energy of the proton quasi-bound state coincides with the experimental
transition energy. In particular, even the most realistic calculations do
not include pairing correlations, or it is just the spectroscopic factor
that is calculated in the simple BCS scheme. In Ref.\cite{BK.89} 
it was shown that the theoretical half-lives calculated using the proton
orbital $1/2^{+}[420]$ and $\beta _{2}\approx 0.05-0.10$ for $^{109}$I, and $%
3/2^{+}[422]$ and $\beta _{2}\approx 0.10-0.15$ for $^{113}$Cs, are
consistent with the experimental values.

The RHB and FRDM models predict
similar values for the one-proton separation energies of 
proton emitters with $Z \geq 59$. Again, the differences
are largest for I and Cs. The RHB predicts stronger binding for I, and
weaker binding for Cs. When compared with available experimental proton
energies, the FRDM model is very close to experiment for $^{109}$I and $%
^{112}$Cs, while the RHB model does better for $^{113}$Cs. However, both
models fail to reproduce the observed anomaly in the one-proton separation
energies of $^{112}$Cs and $^{113}$Cs. The experimental data indicate that $%
^{112}$Cs is more bound, or better, less unbound. This discrepancy is easily
explained. While in the odd-even nucleus $^{112}$Cs the neutrons are all
paired off, $^{113}$Cs has an odd number of protons and an odd number of
neutrons. Since $N-Z$ is only 2, one expects a relatively strong interaction
between the odd neutron and the odd proton. Compared to the odd-even system,
the additional interaction will increase the binding energy. Such an
additional interaction could be represented, for instance, by a surface
delta-force. Since the two mean-field models do not include any residual
proton-neutron interaction, they cannot reproduce the inversion of the
separation energies. The same effect could be also expected for the proton
separation energies of $^{108}$I and $^{109}$I. Ground-state proton emission
has not been observed in $^{108}$I. The nucleus has been found to be an
alpha emitter, with an upper limit for the proton decay branch of $1 \%$.
This limit implies a reduction in the $Q_p$ value of at least 220 keV
compared to $^{109}$I \cite{WD.97}, and the origin of this additional
binding is again the residual interaction between the odd proton and the odd
neutron.

The theoretical separation energies are also compared with recently reported
experimental data on proton radioactivity from $^{131}$Eu, $^{141}$Ho \cite
{Dav.98}, $^{145}$Tm \cite{Bat.98}, $^{146}$Tm \cite{Liv.93}, and $^{147}$Tm 
\cite{Sel.93}. The $^{131}$Eu transition has an energy $E_p = 0.950 (8)$ MeV
and a half-life 26(6) ms, consistent with decay from either $3/2^+[411]$ or $%
5/2^+[413]$ Nilsson orbital. For $^{141}$Ho the transition energy is $E_p =
1.169 (8)$ MeV, and the half-life 4.2(4) ms is assigned to the decay of the $%
7/2^-[523]$ orbital. The calculated RHB proton separation energy, both for $%
^{131}$Eu and $^{141}$Ho, is $-0.9$ MeV. In the RHB calculation for $^{131}$%
Eu the odd proton occupies the $5/2^+[413]$ orbital, while the ground state
of $^{141}$Ho corresponds to the $7/2^-[523]$ proton orbital. This orbital
is also occupied by the odd proton in the calculated ground states of $%
^{145} $Tm, $^{146}$Tm and $^{147}$Tm. For the proton separation energies we
obtain: $-1.43$ MeV in $^{145}$Tm, $-1.20$ MeV in $^{146}$Tm, and $-0.96$
MeV in $^{147}$Tm. These are compared with the experimental values for
transition energies: $E_p = 1.728 (10)$ MeV in $^{145}$Tm, $E_p = 1.120 (10)$
MeV in $^{146}$Tm, and $E_p = 1.054 (19)$ MeV in $^{147}$Tm. When compared
with spherical WKB or DWBA calculations \cite{ASN.97}, the experimental
half-lives for the three Tm isotopes are consistent with spectroscopic
factors for decays from the $h_{11/2}$ proton orbital. Though our predicted
ground-state configuration $7/2^-[523]$ indeed originates from the spherical 
$h_{11/2}$ orbital, we find that these nuclei are deformed. $^{145}$Tm has a
prolate quadrupole deformation $\beta_2 = 0.23$; $^{146}$Tm and $^{147}$Tm
are oblate in the ground-state with $\beta_2 = -0.20$ and $\beta_2 = -0.19$,
respectively. Calculations also predict possible proton emitters $^{136}$Tb
and $^{135}$Tb with separation energies $-0.90$ MeV and $-1.15$ MeV,
respectively. In both isotopes the predicted ground-state proton
configuration is $3/2^+[411]$. Another possible proton emitter is $^{130}$Eu
with separation energy $-1.22$ MeV and the last occupied proton orbital $%
5/2^-[532]$ or $5/2^+[413]$. In our calculation these two levels are almost
degenerate, and the resulting binding and separation energies are identical.

The calculated mass quadrupole deformation parameters for the odd-Z nuclei $%
53\leq Z\leq 69$ at and beyond the drip line are shown in Fig. \ref{figC}.
While prolate deformations $0.15\leq \beta _{2}\leq 0.20$ are calculated for
most of the I and Cs nuclei, the proton-rich isotopes of La, Pr, Pm, Eu and
Tb are strongly prolate deformed ($\beta _{2}\approx 0.30-0.35$). By
increasing the number of neutrons, Ho and Tm display a transition from
prolate to oblate shapes. The absolute values of $\beta _{2}$ decrease as we
approach the spherical solutions at $N=82$. The quadrupole deformations
calculated with the RHB/NL3 effective interaction are found in
excellent agreement with the predictions of the macroscopic-microscopic mass
model \cite{MN.95}. 

A detailed analysis of single proton levels, including spectroscopic
factors, can be performed in the canonical basis which results from the
fully microscopic and self-consistent RHB calculations. For the Eu isotopes
this is illustrated in Fig. \ref{figD}, where we display the proton
single-particle energies in the canonical basis as function of the neutron
number. The dashed line denotes the position of the Fermi level. The proton
energies are the diagonal matrix elements of the Dirac Hamiltonian $h_{D}$
in the canonical basis. The phase-space of positive-energy states should not
be confused with the continuum of scattering states which asymptotically
behave as plane waves. The RHB ground-state wave function can be written
either in the quasiparticle basis as a product of independent quasi-particle
states, or in the {\it canonical basis} as a highly correlated BCS-state. In
the {\it canonical basis} nucleons occupy single-particle states. The
canonical states are eigenstates of the RHB density matrix. The eigenvalues
are the corresponding occupation numbers. In particular, we notice that for
the proton emitter $^{131}$Eu, the ground-state corresponds to the odd
valence proton in the $5/2^{+}[413]$ orbital. For $^{130}$Eu the states $%
5/2^{+}[413]$ and $5/2^{-}[532]$ are almost degenerate, and this is an
example of a situation in which only the comparison of calculated and
measured half-lives can decide which is the last occupied proton orbital.
Both for $^{130}$Eu and $^{131}$Eu the macroscopic-microscopic model
predicts the last proton orbital to be $\Omega ^{\pi }=3/2^{+}$, which
corresponds to the Nilsson orbital $3/2^{+}[411]$. In the energy diagram of
Fig. \ref{figD} the states $5/2^{+}[413]$ and $5/2^{-}[532]$ are closer to
the Fermi level for all calculated Eu isotopes.

In Fig. \ref{figE} we plot the proton single-particle energies in the
canonical basis for the nuclei at the drip-line: $^{118}$La, $^{124}$Pr, $%
^{129}$Pm, $^{134}$Eu, $^{139}$Tb. The levels are shown as functions of the
number of protons, and again the dashed line denotes the position of the
Fermi level. From this diagram one can easily deduce which levels are most
likely to be occupied by the odd-proton, at and beyond the drip-line. Figure 
\ref{figE} should be compared with Table \ref{TabA} for the odd-proton
orbitals occupied in the proton emitters. Of course, by further decreasing
the number of neutrons, i.e. by going beyond the drip-line, the proton
levels and the Fermi level are shifted upwards in energy. However,
as can be seen from Fig. \ref{figD} for the Eu isotopes, this increase in
energy is very smooth and dramatic changes should not be expected for the
occupation of the proton orbitals beyond the drip-line.

In many applications of the HFB and RHB models it has been emphasized 
how important is the self-consistent treatment of pairing correlations in
drip-line nuclei. 
In order to illustrate the relative contribution of the pairing
correlations to the binding energy, for the drip-line nuclei $51\leq Z\leq 69
$, in Fig. \ref{figF} we display the average values of the proton pairing
gaps for occupied canonical states 
\begin{equation}
<\Delta _{p}>={\frac{{\sum_{[\Omega ^{\pi }]}\Delta _{\lbrack \Omega ^{\pi}]}v_{[\Omega ^{\pi }]}^{2}}}{{\sum_{[\Omega ^{\pi }]}v_{[\Omega ^{\pi
}]}^{2}}}},  \label{ang}
\end{equation}
where $v_{[\Omega ^{\pi }]}^{2}$ are the occupation probabilities of the
proton orbitals, and $\Delta _{\lbrack \Omega ^{\pi }]}$ are the diagonal
matrix elements of the pairing part of the RHB single-nucleon Hamiltonian in
the canonical basis. The plotted values of $<\Delta _{p}>$ correspond to the
RHB self-consistent solutions for the ground-states of the drip-line nuclei.
The pairing correlations vanish for Sb and I. The average proton pairing gap
is $1-1.5$ MeV for drip-line nuclei between $^{115}$Cs and $^{139}$Tb, and
it increases to more than 2 MeV for $^{146}$Ho and $^{152}$Tm. 

\bigskip
%
\section{Conclusions}

The relativistic Hartree-Bogoliubov theory has been used to study the 
ground-state properties of deformed 
proton emitters with 53$\leq Z\leq $ 69. The principal advantage of working
in the relativistic framework is its ability to describe all 
the nuclear mean fields in a unified and fully self-consistent way. 
In particular, the spin-orbit term of the effective single-nucleon 
potential is automatically included in the relativistic mean-field model. 
In contrast to non-relativistic models, the strength and isospin dependence
of the spin-orbit interaction do not require the introduction 
of additional free parameters. With the addition of the pairing force,
the relativistic Hartree-Bogoliubov theory provides a unified and
self-consistent description of mean-field and pairing correlations. 
This theoretical framework is especially important for applications
in exotic nuclei far from the valley of $\beta$-stability. The model
parameters are the coupling constants and meson masses of the mean-field
Lagrangian, and the interaction in the $pp$ channel. The NL3 parameter
set has been used for the mean-field Lagrangian, and pairing correlations
have been described by the pairing part of the finite range Gogny 
interaction D1S. This particular combination of effective forces 
in the $ph$ and $pp$ channels has been used in most of our applications
of the RHB theory. The calculated ground-state properties have been
found in excellent agreement with experimental data, not only for 
spherical $\beta$-stable nuclei, but also for nuclear systems with 
large isospin values on both sides of the valley of $\beta$-stability.
No attempt has been made to adjust the model parameters to the specific
properties of nuclei studied in this work.

The model has been used to study the location of the proton drip-line, 
the ground-state quadrupole deformations and one-proton separation 
energies at and beyond the drip line, the deformed single particle 
orbitals occupied by the odd valence proton, and the corresponding 
spectroscopic factors. The results of fully self-consistent calculations
have been compared with predictions of the finite range droplet mass
model, and with very recent experimental data on ground-state proton
emitters. 

The calculated ground-state mass deformation parameters have been found
in excellent agreement with values predicted by the macroscopic-microscopic 
FRDM model calculations. In particular, both models
predict a transition from prolate to oblate shapes in the Tm isotopes. 
In many cases, however, different single particle orbitals occupied by
the odd proton are predicted by the two models. Correspondingly, in these
cases relatively large differences for the one-proton separation energies
have been calculated. These divergences might be related to the different
descriptions of the effective spin-orbit potential. In comparison with 
recent experimental data on ground-state proton emitters, the calculated 
one-proton separation energies are in excellent agreement with 
transitions energies in $^{113}$Cs, $^{131}$Eu, $^{141}$Ho, $^{146}$Tm, 
and $^{147}$Tm. For $^{145}$Tm, however, the calculated and experimental
proton energies differ by almost 300 keV. Both the RHB and FRDM models
fail to reproduce the observed anomaly in the proton energies of 
the ground state emitters $^{112}$Cs and $^{113}$Cs. The anomaly can
probably be related to the short range proton-neutron interaction, 
not included in the mean-field models. Summarizing, we have found 
a remarkable agreement between results of RHB calculations and 
very recent experimental data on deformed ground-state proton
emitters. It should be emphasized that these results have been
obtained with the standard RHB parameterization, i.e. no new parameters
have been introduced to specifically describe the proton emitters. 
In addition, the spectroscopic factors which correspond to the 
fully self-consistent RHB solutions will be useful in calculations
of proton partial decay half-lives.  

\newpage
\leftline{\bf ACKNOWLEDGMENTS}

This work has been supported in part by the Bundesministerium f\"{u}r
Bildung und Forschung under project 06 TM 875. 
One of the authors (GAL) acknowledges with thanks the warm 
hospitality of the National Superconducting Cyclotron Laboratory
at MSU during his stay there.

\newpage

\begin{figure}
\caption{Predictions of the RHB model (NL3 $+$ D1S) for the one-proton
separation energies of odd-Z nuclei $61 \leq Z \leq 69$, at and beyond the
drip-line.}
\label{figA}
\end{figure}

\begin{figure}
\caption{
Same as in Fig. \ref{figA}, but for odd-Z isotopes $53 \leq Z \leq
59$.}
\label{figB}
\end{figure}

\begin{figure}
\caption{
Self-consistent ground-state quadrupole deformations for odd-Z
nuclei $53 \leq Z \leq 69$, at and beyond the proton drip-line.}
\label{figC}
\end{figure}

\begin{figure}
\caption{
The proton single-particle levels for the Eu isotopes. The dashed
line denotes the position of the Fermi level. The energies in the canonical
basis correspond to ground-state solutions calculated with the RHB model
(NL3 $+$ D1S).}
\label{figD}
\end{figure}

\begin{figure}
\caption{
The proton single-particle levels for the isotopes at the proton
drip-line $57 \leq Z \leq 65$. The energies in the canonical basis are
plotted as functions of the proton number Z, and the dashed line denotes the
position of the Fermi level.}
\label{figE}
\end{figure}

\begin{figure}
\caption{
Average proton pairing gaps $<\Delta _{p}>$ for the isotopes at
the proton drip-line: $^{110}$I, $^{115}$Cs, $^{118}$La, $^{124}$Pr,
$^{129}$Pm, $^{134}$Eu, $^{139}$Tb, $^{146}$Ho, and $^{152}$Tm.}
\label{figF}
\end{figure}

\newpage
\begin{table}
\caption{ Odd-Z ground-state proton emitters in the region of nuclei with 
$53\leq Z \leq 69$. Results of the RHB calculation for the one-proton
Separation energies $S_p$, quadrupole deformations $\protect\beta_2$, and
the deformed single-particle orbitals occupied by the odd valence proton,
are compared with predictions of the macroscopic-microscopic mass model, and
with the experimental transition energies. All energies are in units of MeV;
the RHB spectroscopic factors are displayed in the sixth column.}
\label{TabA}
\begin{center}
\begin{tabular}{llllllllll}
& N & $S_p$ & $\beta_2$ & $p$-orbital & $u^2$ & $\Omega^{\pi}_p$ \cite
{MNK.97} & $S_p$ \cite{MNK.97} & $\beta_2$ \cite{MN.95} & $E_p$ exp. \\ 
\hline
$^{107}$I & 54 & -1.40 & 0.15 & $3/2^+[422]$ & 0.84 & $1/2^+$ & -2.14 & 0.14
&  \\ 
$^{108}$I & 55 & -0.73 & 0.16 & $3/2^+[422]$ & 0.79 & $1/2^+$ & -1.12 & 0.15
&  \\ 
$^{109}$I & 56 & -0.37 & 0.16 & $3/2^+[422]$ & 0.81 & $1/2^+$ & -0.95 & 0.16
& 0.8126(40) \cite{Sel.93} \\ 
$^{111}$Cs & 56 & -1.97 & 0.20 & $1/2^+[420]$ & 0.74 & $3/2^+$ & -1.43 & 0.19
&  \\ 
$^{112}$Cs & 57 & -1.46 & 0.20 & $1/2^+[420]$ & 0.74 & $3/2^+$ & -0.76 & 0.21
& 0.807(7) \cite{Page.94} \\ 
$^{113}$Cs & 58 & -0.94 & 0.21 & $1/2^+[420]$ & 0.73 & $3/2^+$ & -0.76 & 0.21
& 0.9593(37) \cite{Bat.98} \\ 
$^{115}$La & 58 & -1.97 & 0.26 & $1/2^+[420]$ & 0.20 & $3/2^+$ & -1.50 & 0.27
&  \\ 
$^{116}$La & 59 & -1.09 & 0.30 & $3/2^-[541]$ & 0.73 & $3/2^+$ & -0.67 & 0.28
&  \\ 
$^{119}$Pr & 60 & -1.40 & 0.32 & $3/2^-[541]$ & 0.39 & $3/2^-$ & -1.41 & 0.31
&  \\ 
$^{120}$Pr & 61 & -1.17 & 0.33 & $3/2^-[541]$ & 0.33 & $3/2^-$ & -0.66 & 0.32
&  \\ 
$^{124}$Pm & 63 & -1.00 & 0.35 & $5/2^-[532]$ & 0.72 & $5/2^-$ & -1.34 & 0.33
&  \\ 
$^{125}$Pm & 64 & -0.81 & 0.35 & $5/2^-[532]$ & 0.74 & $5/2^-$ & -1.24 & 0.33
&  \\ 
$^{130}$Eu & 67 & -1.22 & 0.34 & $5/2^-[532]$ & 0.44 & $3/2^+$ & -1.17 & 0.33
&  \\ 
$^{131}$Eu & 68 & -0.90 & 0.35 & $5/2^+[413]$ & 0.44 & $3/2^+$ & -1.01 & 0.33
& 0.950(8) \cite{Dav.98} \\ 
$^{135}$Tb & 70 & -1.15 & 0.34 & $3/2^+[411]$ & 0.62 & $3/2^+$ & -1.15 & 0.33
&  \\ 
$^{136}$Tb & 71 & -0.90 & 0.32 & $3/2^+[411]$ & 0.65 & $3/2^+$ & -0.55 & 0.31
&  \\ 
$^{140}$Ho & 73 & -1.10 & 0.31 & $7/2^-[523]$ & 0.61 & $7/2^-$ & -0.81 & 0.30
&  \\ 
$^{141}$Ho & 74 & -0.90 & 0.32 & $7/2^-[523]$ & 0.64 & $7/2^-$ & -0.89 & 0.29
& 1.169(8) \cite{Dav.98} \\ 
$^{145}$Tm & 76 & -1.43 & 0.23 & $7/2^-[523]$ & 0.47 & $1/2^+$ & -1.0 & 0.25
& 1.728(10) \cite{Bat.98} \\ 
$^{146}$Tm & 77 & -1.20 & -0.21 & $7/2^-[523]$ & 0.50 & $7/2^-$ & -0.60 & 
-0.20 & 1.120(10) \cite{Liv.93} \\ 
$^{147}$Tm & 78 & -0.96 & -0.19 & $7/2^-[523]$ & 0.55 & $7/2^-$ & -0.56 & 
-0.19 & 1.054(19) \cite{Sel.93}
\end{tabular}
\end{center}
\end{table}

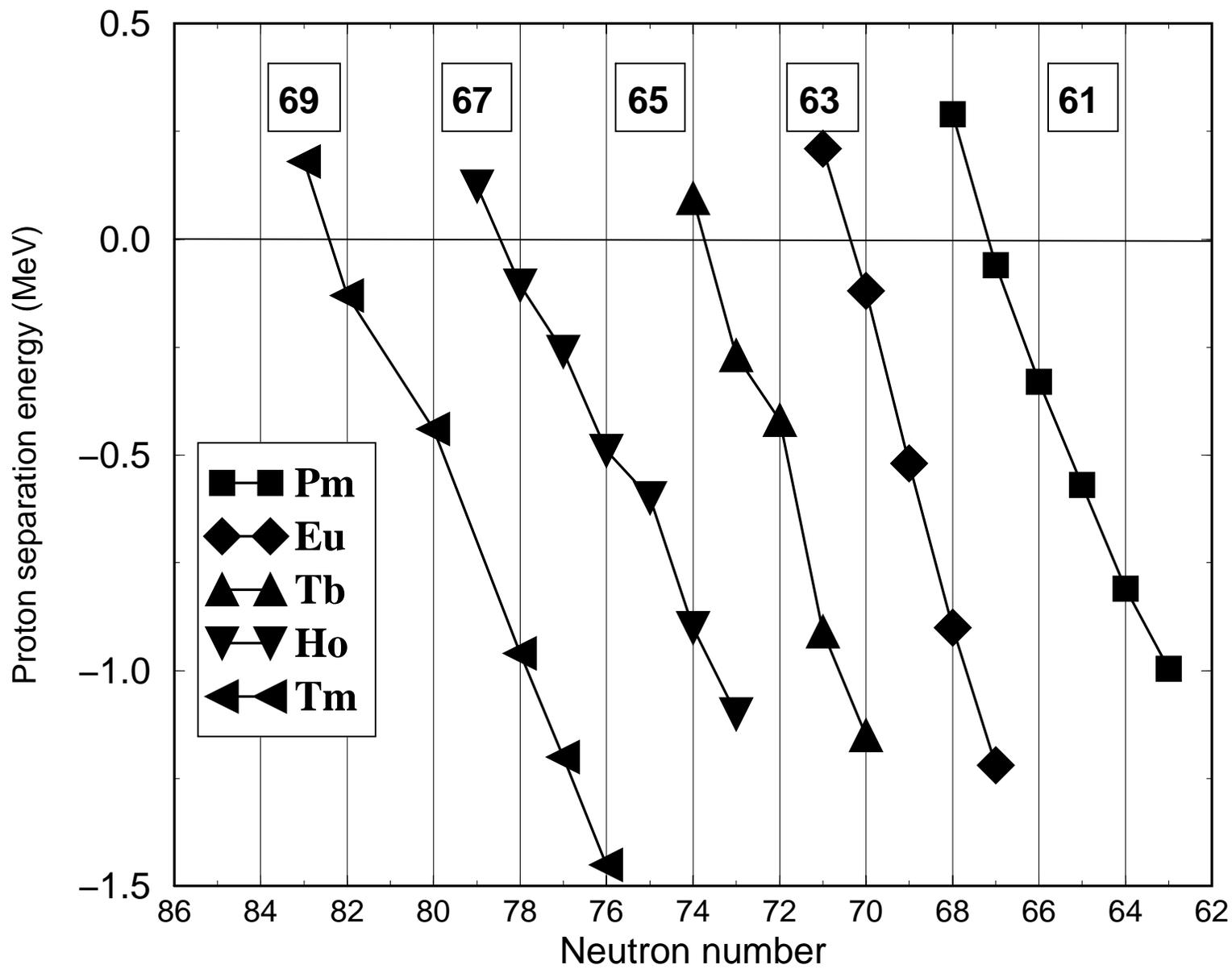
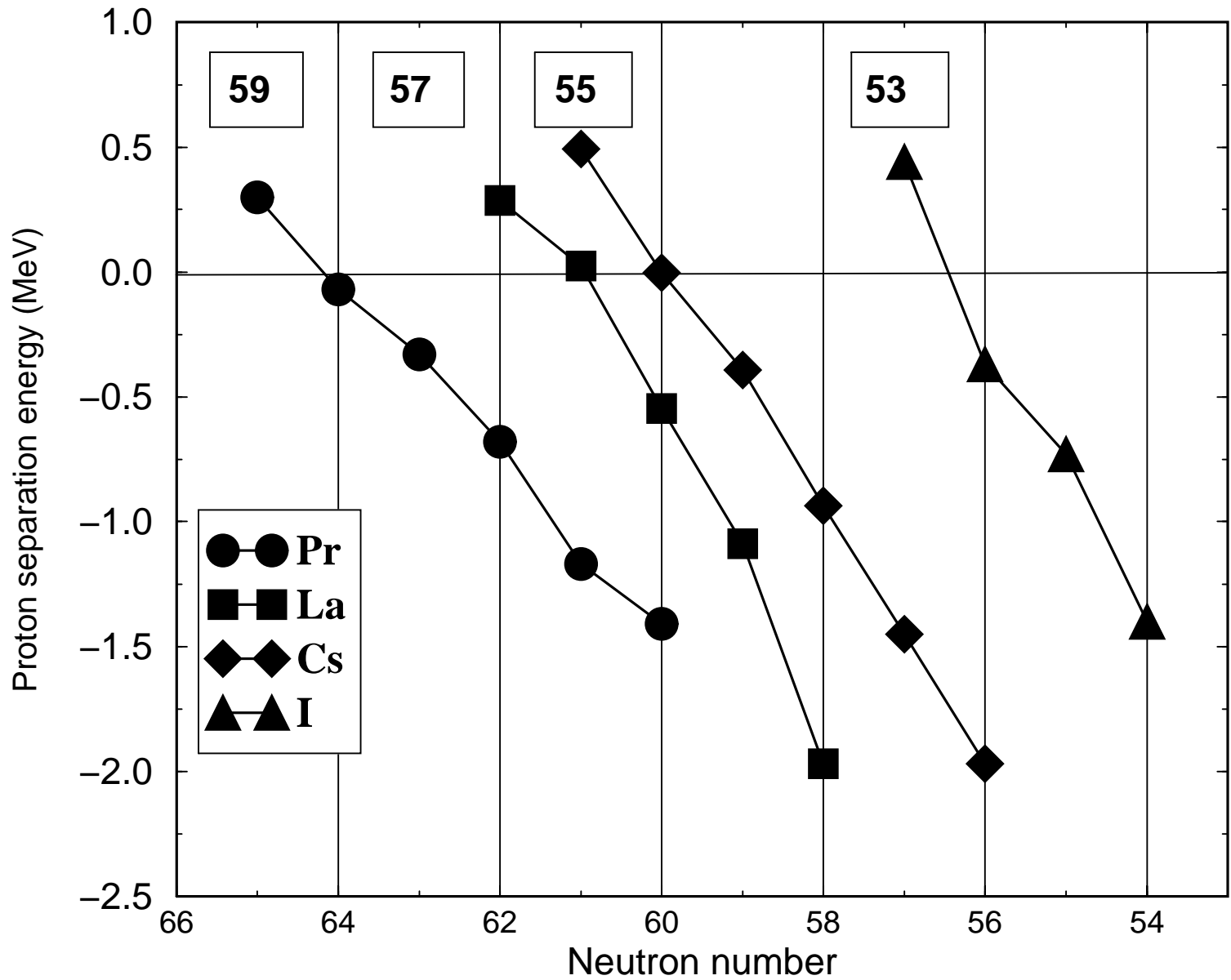
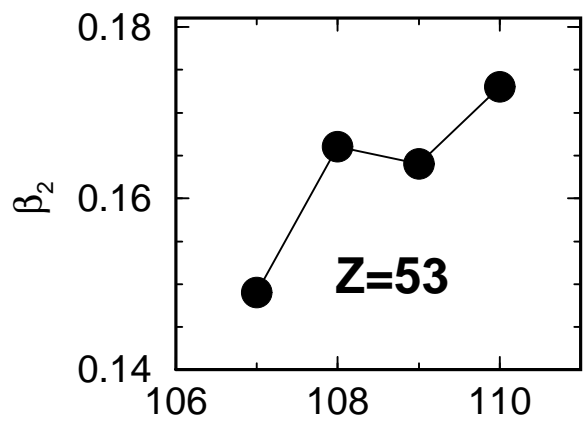
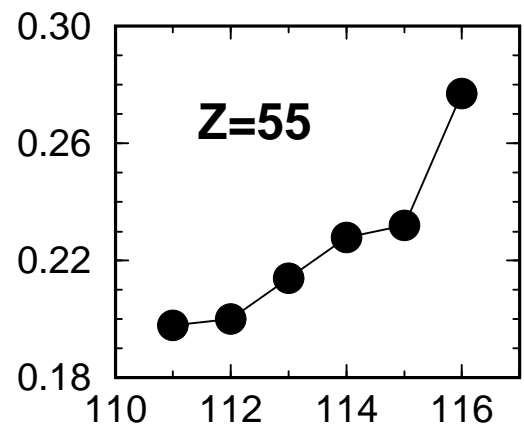
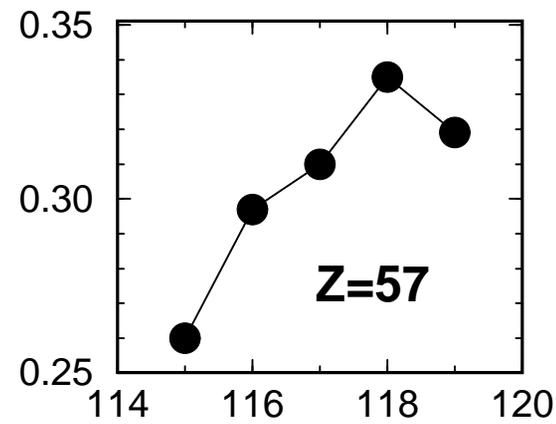
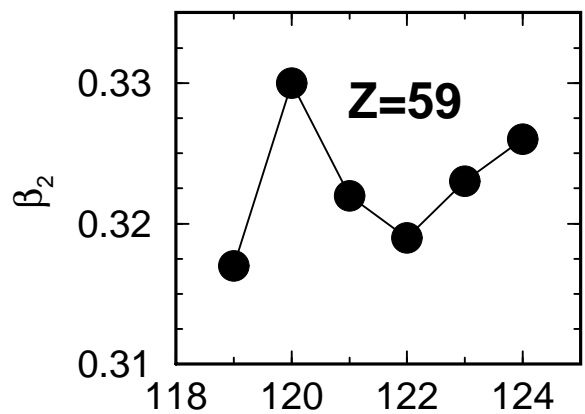
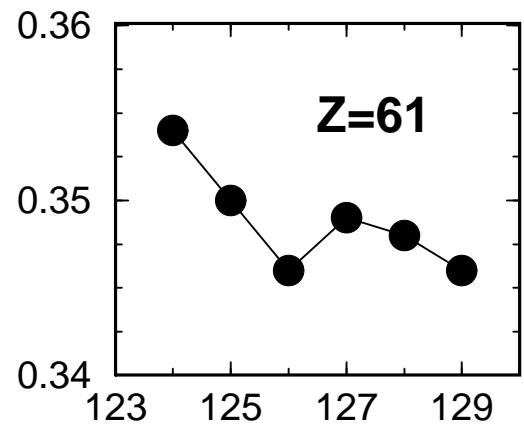
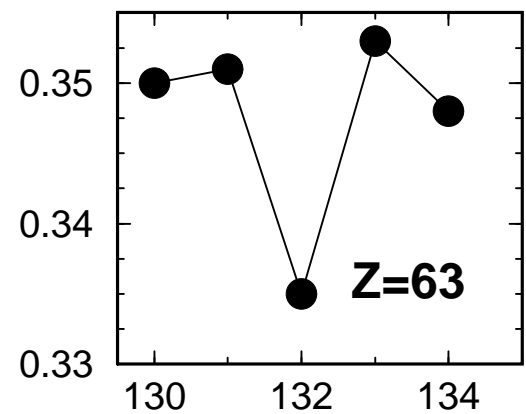
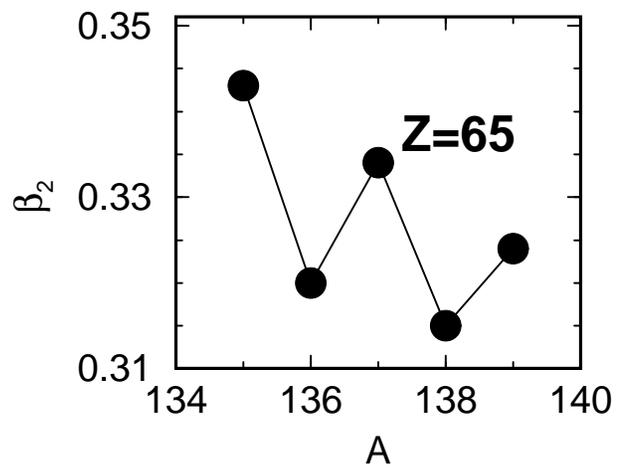
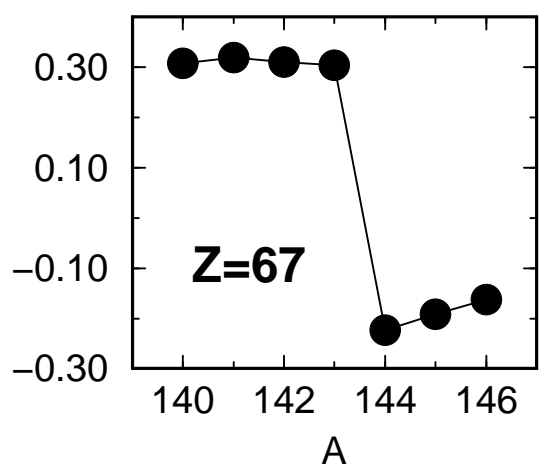
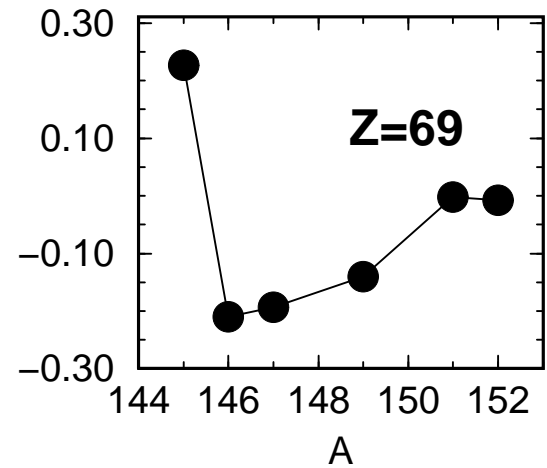
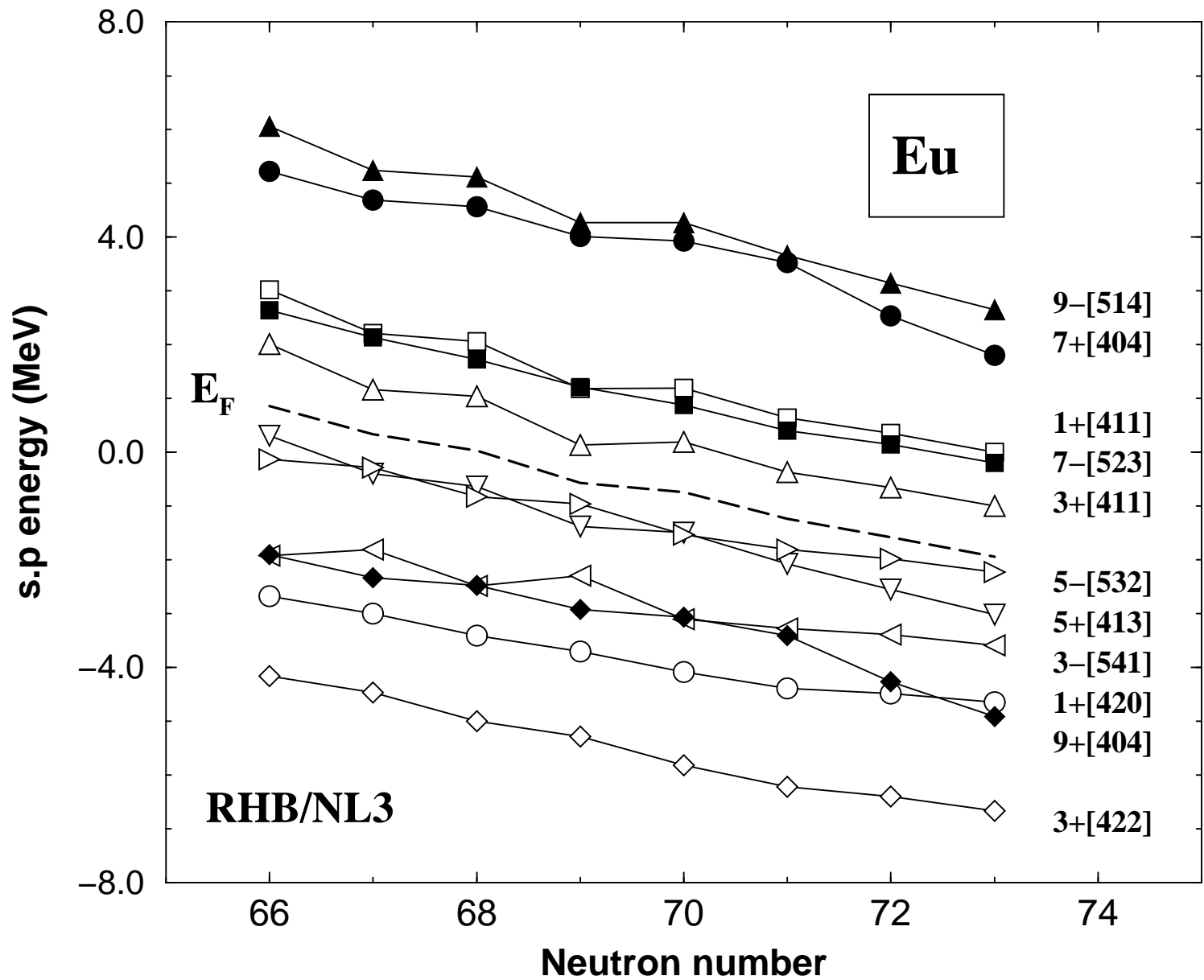
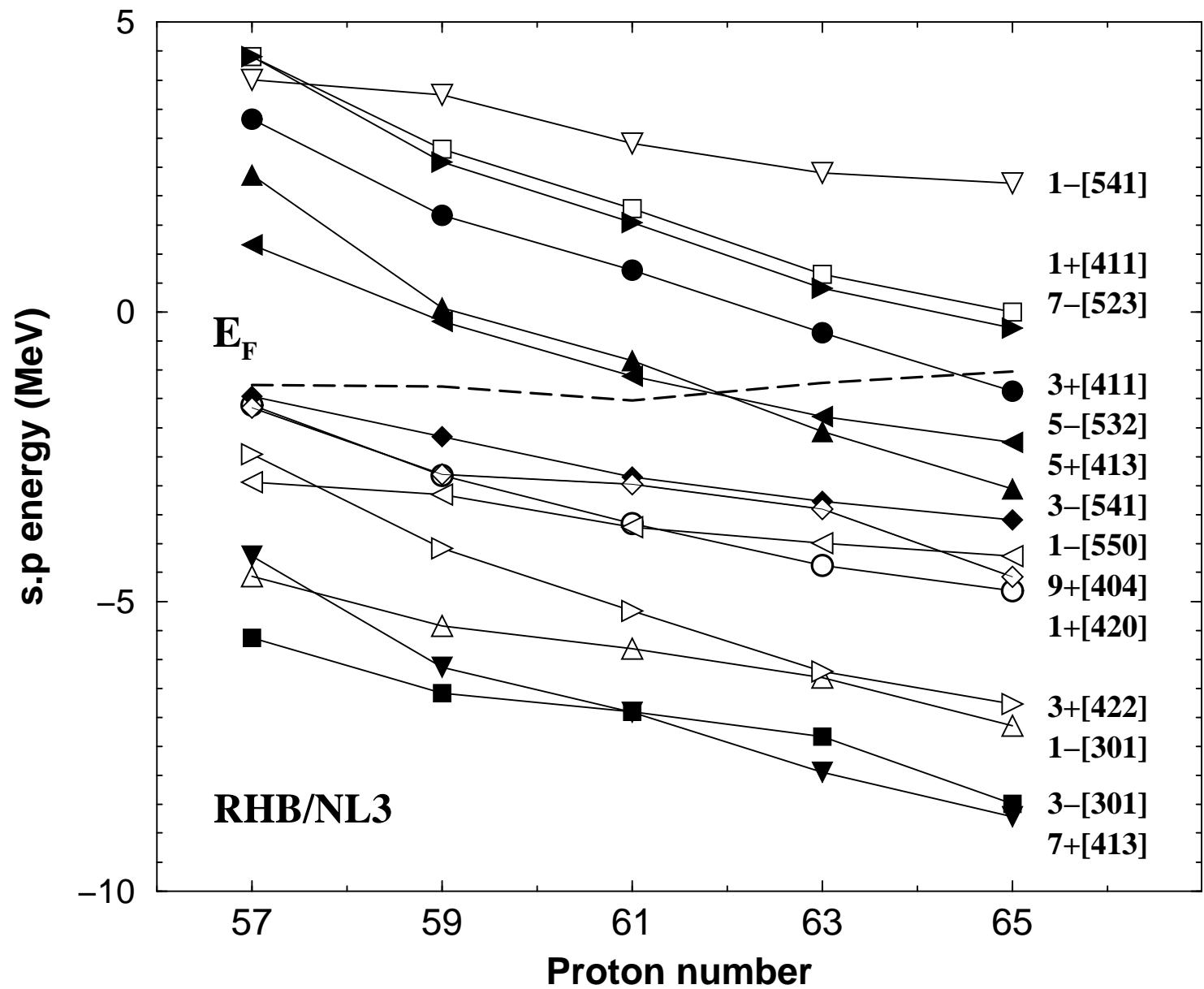
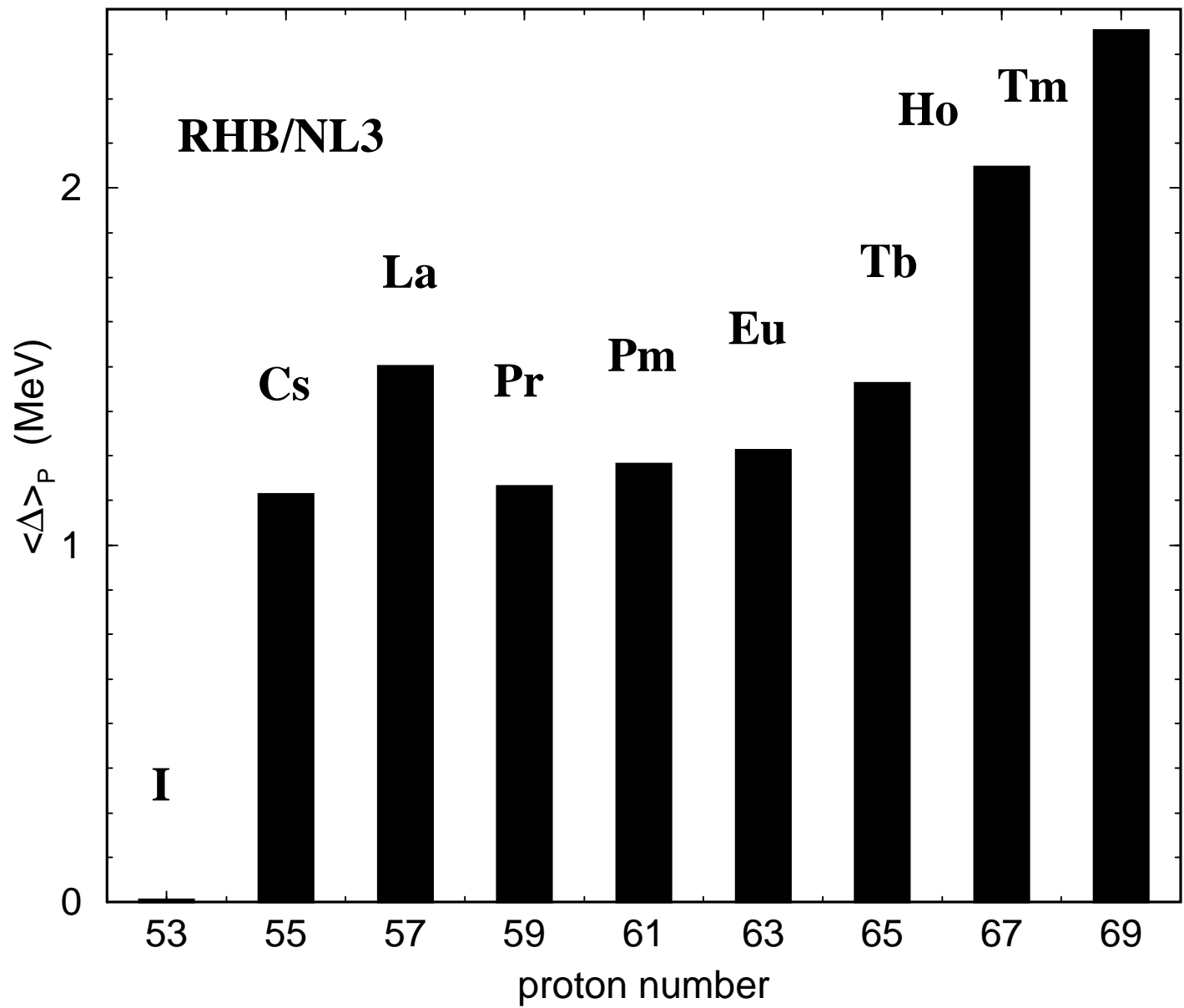
\end{document}